\documentclass[preprint,onecolumn,nofootinbib]{revtex4}
\pdfoutput=1
\usepackage[colorlinks=true,linkcolor=blue,urlcolor=blue,filecolor=black,citecolor=red,pdfstartview=FitV,pdftitle={},pdfsubject={},pdfkeywords={},pdfpagemode=None,bookmarksopen=true]{hyperref}
\usepackage{graphicx}
\usepackage{amsmath}
\usepackage{amsfonts}
\usepackage{amssymb,ulem}
\usepackage{color,xcolor}%
\usepackage{CJK}
\usepackage{subfigure}
\usepackage{appendix}

\usepackage{dcolumn}
\usepackage{fmtcount}

\newcommand{\f}{\begin{equation}}
\newcommand{\ff}{\end{equation}}
\newcommand{\fa}{\begin{eqnarray}}
\newcommand{\ffa}{\end{eqnarray}}

\begin{document}
\title{Chaotic Dynamics around a Hairy Black Hole in Horndeski Gravity}

\author{Yang Yu$^{1}$}
\thanks{mx120230358@stu.yzu.edu.cn}
\author{Ruo-Ting Chen$^{1}$}
\thanks{ruotingchen@163.com}
\author{Shulan Li$^{2}$}
\thanks{shulanli.yzu@gmail.com}
\author{Dan Zhang$^{1}$}
\thanks{danzhanglnk@163.com}
\author{Jian-Pin Wu$^{1}$}
\thanks{jianpinwu@yzu.edu.cn, corresponding author}

\affiliation{
$^{1}$ \mbox{Center for Gravitation and Cosmology, College of Physical Science and Technology,} \mbox{Yangzhou University, Yangzhou 225009, China}\\
$^{2}$ \mbox{Department of Physics, Shanghai University, Shanghai 200444, China}
}

\begin{abstract}

We investigate the chaotic dynamics of a spinning test particle in a static, spherically symmetric hairy black hole in Horndeski gravity, using the Mathisson-Papapetrou-Dixon equations closed by the Tulczyjew-Dixon condition. Bound orbits are identified by fixing the angular momentum and setting the energy according to the effective potential, so that the accessible parameter range is bounded by a critical hair parameter, $h_{\rm crit}$, which varies non-monotonically with the spin. We analyze the resulting chaotic behavior using Lyapunov exponents, Poincaré sections, and reconstructed attractors. We find that the response of these indicators to scalar hair is orbit- and spin-dependent rather than universally monotonic. An analysis of orbital frequency ratios shows that, at physically maximal spin ($S=1$), an approaching $2{:}1$ azimuthal resonance overlapping a persistent $1{:}1$ polar lock amplifies chaos, while the hair-driven erosion of potential saddles tends to suppress it. At non-physical spin ($S=1.5$), the Lyapunov exponent decreases monotonically. Gravitational waveforms constructed via the quadrupole formula inherit the same state-specific pattern: the correlation between the two polarization modes varies non-monotonically with $h$, in step with the Lyapunov exponent. These results show that chaotic indicators respond to scalar hair in an orbit- and resonance-dependent manner rather than through simple monotonic suppression, and identify the spin-dependent resonance structure as an essential ingredient for interpreting chaos around hairy black holes.

\end{abstract}

\maketitle
\tableofcontents

\section{Introduction}
The “no-hair theorem'', formulated in the early 1970s, posits that an isolated, stationary black hole (BH) in general relativity (GR) is uniquely characterized by only three classical parameters: mass, electric charge, and angular momentum~\cite{Israel:1967wq, Israel:1967za, Carter:1971zc, Wald:1971iw}. This profound result implies that all other information about the progenitor matter that formed the BH or fell into it—the “hair”—is irretrievably lost behind the event horizon. While this theorem holds rigorously in vacuum Einstein-Maxwell theory, its validity in more general gravitational frameworks remains a subject of intense investigation. Numerous counterexamples and extensions have been explored, particularly in theories involving scalar fields, leading to the concept of “hairy” BHs. A particularly fertile framework for such solutions is Horndeski gravity, the most general scalar-tensor theory with second-order equations of motion. Horndeski theories not only provide a consistent platform for infrared modifications of gravity, potentially addressing late-time cosmic acceleration, but also admit exact, asymptotically flat BH solutions endowed with nontrivial scalar hair~\cite{Kase:2018aps}.

Despite the theoretical appeal of hairy BHs, any viable modified gravity theory faces stringent empirical constraints. A wealth of experiments—from solar system tests~\cite{Will:2014kxa} to binary pulsar dynamics and gravitational wave (GW) observations~\cite{LIGOScientific:2016lio,LIGOScientific:2016sjg,LIGOScientific:2019fpa}—have consistently confirmed GR’s remarkable accuracy, particularly within the well-tested weak-field regime. Even in the strong-field regime, deviations from GR, such as those induced by scalar hair, are expected to be exceedingly faint. Detecting these subtle signatures therefore demands exceptionally sensitive probes.

Chaotic dynamical systems, renowned for their sensitivity to initial conditions and parameter variations, offer a promising avenue for such high-precision tests. In GR, the integrability of geodesic motion can be broken by additional degrees of freedom, such as particle spin. As demonstrated by Suzuki and Maeda~\cite{Suzuki:1996gm}, a spinning test particle in Schwarzschild spacetime can exhibit chaotic motion beyond a critical spin, governed by the Mathisson-Papapetrou-Dixon (MPD) equations~\cite{Mathisson:1937zz,Papapetrou:1951pa,Dixon:1970zza}. This chaos originates from saddle points in the effective potential, leading to heteroclinic orbits and is quantifiable via positive Lyapunov exponents (LEs) and broken tori in Poincaré sections. The orbital dynamics of stars or compact objects near supermassive BHs, such as the S-stars orbiting Sgr A*, thus constitute a natural laboratory where minute metric deviations could be amplified through chaotic evolution, imprinting on both trajectories and GW emissions.

Recent advances in observational techniques—from the optical imaging by the Event Horizon Telescope~\cite{EventHorizonTelescope:2019dse,EventHorizonTelescope:2019ggy,EventHorizonTelescope:2019jan,EventHorizonTelescope:2019pgp,EventHorizonTelescope:2019ths,EventHorizonTelescope:2019uob,EventHorizonTelescope:2022apq,EventHorizonTelescope:2022exc,EventHorizonTelescope:2022urf,EventHorizonTelescope:2022wok,EventHorizonTelescope:2022xqj} to the detection of GWs~\cite{LIGOScientific:2016sjg,LIGOScientific:2016emj,LIGOScientific:2016lio,LIGOScientific:2016vbw,LIGOScientific:2016vlm,LIGOScientific:2016vpg,LIGOScientific:2017bnn,LIGOScientific:2018mvr,LIGOScientific:2019fpa}—have opened new windows onto strong-field gravity. Concurrently, efforts to constrain or detect BH hair through these channels are actively underway~\cite{Isi:2019aib,Johannsen:2010ru,Johannsen:2011dh,Khodadi:2020jij,Broderick:2013rlq, Vagnozzi:2022moj}. Chaotic dynamics offers another window onto such deviations. Since chaos emerges from unstable structures in the effective potential, its properties are sensitive to the detailed geometry modified by the scalar hair.

In this work, we study the chaotic dynamics of a spinning test particle, governed by the MPD equations, in the hairy BH spacetime of Horndeski gravity. Our primary goal is to investigate how the scalar hair influences the onset and characteristics of the chaos. As we shall see, this influence is not simply monotonic: the chaotic indicators respond to the hair in a state-specific waythat is largely governed by resonances between the orbital degrees of freedom. We compute LEs and Poincaré maps across a range of hairy and spin parameters, quantify the resonance content of the motion, and use the quadrupole formula to construct and analyze the gravitational waveform emitted by the chaotic orbiting particle, examining the modifications induced by the scalar hair.

This paper is organized as follows. Sec.~\ref{sec-1} introduces the Horndeski gravity framework and hairy BH solutions, derives the MPD equations for spinning test particles, specifies the initial conditions, and numerically analyzes the characteristics of the resulting orbital motion. Sec.~\ref{sec-2} investigates the effective potential, classifying its types and elucidating the role of saddle points in mediating chaotic behavior. Sec.~\ref{sec-3} analyzes chaotic dynamics, employing Takens’ embedding theory for phase space reconstruction, the Wolf algorithm to compute LEs, and Poincaré sections together with a resonance-phase analysis to characterize the state-specific response of the chaos to scalar hair. Sec.~\ref{sec-4} extends the analysis to GW signatures, constructing waveforms via the quadrupole formula and quantifying phase coherence between polarization modes. Finally, Sec.~\ref{conclusion} summarizes the key findings and suggests future research directions.

\section{Spinning Particle Dynamics in a Hairy Spacetime}\label{sec-1}

In this section, we investigate the orbital dynamics governed by the MPD equations in the hairy BH spacetime of Horndeski gravity. A brief review of the hairy BH solution in this theory is provided in subsection~\ref{sub-1-1}. In subsection~\ref{sub-1-2}, we derive the corresponding MPD equations that govern the motion of a spinning test particle, and in subsection~\ref{sub-1-3} we specify the necessary initial conditions. Finally, we numerically integrate the MPD equations in the hairy BH background and analyze the characteristics of the resulting orbital motion in subsection~\ref{sub-1-4}.

\subsection{Horndeski Theory and the Hairy BH Solution}\label{sub-1-1}

We consider the most general scalar-tensor theory yielding second-order equations of motion in 4 dimensions, namely the Horndeski theory. Its action can be written as~\cite{Babichev:2017guv, Bergliaffa:2021diw}
\begin{equation}\label{eq:action}
    \begin{aligned}
        S &= \int \mathrm{d}^{4}x \sqrt{-g} [ Q_2 + Q_3\Box\varphi + Q_{4}R +Q_{4,\chi}( (\Box\varphi)^2 - (\nabla^{\mu}\nabla^{\nu}\varphi)(\nabla_{\mu}\nabla_{\nu}\varphi) ) + Q_{5}G_{\mu\nu}\nabla^{\mu}\nabla^{\nu}\varphi \\
        &\quad - \frac{1}{6}Q_{5,\chi}( (\Box\varphi)^{3} - 3(\Box\varphi)(\nabla^{\mu}\nabla^{\nu}\varphi)(\nabla_{\mu}\nabla_{\nu}\varphi) + 2(\nabla_{\mu}\nabla_{\nu}\varphi)(\nabla^{\nu}\nabla^{\gamma}\varphi)(\nabla_{\gamma}\nabla^{\mu}\varphi) ) ],
    \end{aligned}
\end{equation}
where $\varphi$ is the scalar field, $\chi = -\frac{1}{2}\partial^{\mu}\varphi\partial_{\mu}\varphi$ denotes its kinetic term, $Q_{i}\ (i=2,3,4,5)$ are arbitrary functions of $\chi$, $R$ and $G_{\mu\nu}$ are the Ricci scalar and the Einstein tensor respectively, and $Q_{i,\chi}$ stands for the derivative of $Q_i$ with respect to $\chi$~\cite{Babichev:2017guv}. This theory encompasses a broad class of modified gravity models while preserving second-order equations of motion, which is a sufficient condition to avoid Ostrogradsky ghosts.

In this work, we build on the framework of Ref.~\cite{Bergliaffa:2021diw}, which focuses on a specific subclass of quartic Horndeski theory. The action is given by
\begin{equation}
        S=\int \mathrm{d}^{4}x \sqrt{-g}\left\{Q_{2}(\chi)+Q_{3}(\chi) \square \varphi+Q_{4}(\chi) R+Q_{4}, \chi\left[(\square \varphi)^{2}-\left(\nabla^{\mu} \nabla^{\nu} \varphi\right)\left(\nabla_{\mu} \nabla_{\nu} \varphi\right)\right]\right\} \,,
\end{equation}
with the specific choice of the functions $Q_i$:
\begin{subequations}
    \begin{align}
        Q_2 &= \alpha_{21}\chi + \alpha_{22}(-\chi)^{\omega_2}, \\
        Q_3 &= \alpha_{31}(-\chi)^{\omega_3}, \\
        Q_4 &= (8{\pi}G)^{-2} + \alpha_{42}(-\chi)^{\omega_4}, \\
        Q_5 &= 0,
    \end{align}
\end{subequations}
where $\alpha_{21}=\alpha_{31}=0$, $\omega_{2}=\frac{3}{2}$, and $\omega_4=\frac{1}{2}$. In this case, the field equations admit static, spherically symmetric BH solutions endowed with scalar hair. Following Ref.~\cite{Bergliaffa:2021diw}, we work with geometric units $G=c=1$, and the solution in Schwarzschild coordinates is given by
\begin{subequations}\label{eq:metric}
    \begin{align}
        \mathrm{d}s^{2} &= -f(r)\mathrm{d}t^{2} + f^{-1}(r)\mathrm{d}r^{2} + r^{2}\mathrm{d}\Omega^{2}, \\
        f(r) &= 1 - \frac{2M}{r} - \frac{h}{r}\ln{\frac{r}{2M}}, \\
        \mathrm{d}\Omega^{2} &= \mathrm{d}\theta^{2} + \sin^{2}\theta \mathrm{d}\phi^{2} ,
    \end{align}
\end{subequations}
where $h$ is a parameter that encodes the imprint of the scalar field outside the horizon, i.e., the ``hair''. For $h=0$, the metric~\eqref{eq:metric} reduces to the standard Schwarzschild solution. This family of solutions evades the classical no-hair theorem by relaxing some of its underlying assumptions. In this particular subclass of quartic Horndeski theory, the scalar field configuration is obtained by considering a non-minimal coupling between the scalar field and curvature $(Q_4 R)$, thereby allowing for a nontrivial scalar configuration outside the event horizon. Consequently, the metric~\eqref{eq:metric} provides a self-consistent, analytically tractable background for probing strong-field gravitational effects beyond GR. To simplify the analysis, we henceforth set the BH mass $M=1$, which serves as our fundamental scale.

This solution has been extensively studied in several works, including investigations into the optical appearance of BHs in Ref.~\cite{Wang:2023vcv}, the propagation of external fields in Ref.~\cite{Yang:2023lcm}, and strong gravitational lensing effects in Ref.~\cite{Kumar:2021cyl}. In parallel, the rotating form of Eq.~\eqref{eq:metric}, derived via the Newman-Janis algorithm (NJA), has also been widely examined in Refs.~\cite{Walia:2021emv, Afrin:2021wlj, Lin:2023rmo, Jha:2022tdl, Ghosh:2022kit}.

\subsection{MPD Equations}\label{sub-1-2}
The dynamics of a massive spinning particle in a curved spacetime is governed by MPD equations~\cite{Mathisson:1937zz, Papapetrou:1951pa, Dixon:1970zza}. In covariant form, they read
\begin{subequations}\label{eq:eom}
    \begin{align}
        u^{\mu} &= \frac{\mathrm{d}x^{\mu}}{\mathrm{d}\tau}, \label{eq:eom_x} \\
        \frac{\mathrm{D} p^{\mu}}{\mathrm{d}\tau} &= -\frac{1}{2}{R^{\mu}}_{\nu\alpha\beta}u^{\nu}S^{\alpha\beta}, \label{eq:eom_p} \\
        \frac{\mathrm{D} S^{\mu\nu}}{\mathrm{d}\tau} &= p^{\mu}u^{\nu} - p^{\nu}u^{\mu}, \label{eq:eom_s}
    \end{align}
\end{subequations}
where $\tau$ is an affine parameter along the particle’s worldline, $u^{\mu}$ is the 4‑velocity, $p^{\mu}$ is the generalized 4‑momentum, and $S^{\mu\nu}=-S^{\nu\mu}$ is the antisymmetric spin tensor. The curvature tensor of the background spacetime is denoted by ${R^{\mu}}_{\nu\rho\sigma}$. In the following calculations, we denote the derivatives of the affine parameter as a dot above the variables.

The system~\eqref{eq:eom} is not closed because the relation between $u^{\mu}$ and $p^{\mu}$ is not specified. To close it one must impose a supplementary spin condition (SSC). Different SSCs correspond to different choices of the representative worldline that describes the center of mass~\cite{Witzany:2018ahb}. In this work, we adopt the Tulczyjew-Dixon (TD) condition (known as TD-SSC), which provides a unique worldline~\cite{tulczyjew1959motion, Dixon:1970zza}
\begin{equation}\label{eq:SSC}
    p_{\mu}S^{\mu\nu} = 0.
\end{equation}
Due to TD-SSC, we can define the conserved mass of the particle $m$ as: 
\begin{equation}\label{eq:mass}
    m^2 = -p_{\mu}p^{\mu}.
\end{equation}

Following the procedure in Ref.~\cite{Semerak:1999qc,Tod:1976ud,Han:2016cdh}, we can express the 4‑velocity in terms of $p^{\mu}$ and $S^{\mu\nu}$. To this end, we adopt the normalization convention
\begin{equation}
    p\cdot u \equiv p_{\mu}u^{\mu} = -m,
    \label{eq:pu}
\end{equation}
which fixes the affine parameter $\tau$ and does not follow from the TD-SSC~\eqref{eq:SSC}; see, e.g., Refs.~\cite{Tod:1976ud,Han:2016cdh}. As a consequence of this convention, $\tau$ is not the proper time along the worldline.
Contracting Eq.~\eqref{eq:eom_s} with $p_{\nu}$ gives
\begin{equation}
    p_{\nu}\dot{S}^{\mu\nu}
    = p_{\nu}p^{\mu}u^{\nu}-p_{\nu}p^{\nu}u^{\mu}
    = p^{\mu}(p\cdot u)+m^{2}u^{\mu},
\end{equation}
where we used Eq.\eqref{eq:mass}.
With Eq.\eqref{eq:pu}, this yields:
\begin{equation}
		p_{\nu}{\dot{S}}^{\mu\nu} = m^2u^{\mu}-mp^{\mu}, \label{eq:ps}
\end{equation}

Likewise, lowering the indices in Eq.~\eqref{eq:eom_s} to $\dot{S}_{\mu\nu}=p_{\mu}u_{\nu}-p_{\nu}u_{\mu}$ and contracting with $p^{\mu}$ gives
\begin{equation}
    p^{\mu}\dot{S}_{\nu\mu}
    = p_{\nu}(p\cdot u)+m^{2}u_{\nu}
    = -mp_{\nu}+m^{2}u_{\nu},
\end{equation}
where we used the antisymmetry $\dot{S}_{\nu\mu}=-\dot{S}_{\mu\nu}$.
Solving for $u_{\nu}$ yields:
\begin{equation}
        u^{\nu} = m^{-2}(p_{\mu}{\dot{S}}^{\nu\mu}+mp^{\nu}).\label{eq:u}
\end{equation}
Substituting Eq.~\eqref{eq:u} into ~\eqref{eq:eom_p} and multiplying by $S^{\rho\mu}$ yields:
\begin{equation}
    2m^2\dot{p}_{\mu}S^{\rho\mu} = -mR_{\mu\nu\alpha\beta}p^{\nu}S^{\rho\mu}S^{\alpha\beta}-R_{\mu\nu\alpha\beta}p_{\delta}\dot{S}^{\nu\delta}S^{\rho\mu}S^{\alpha\beta}.
\end{equation}
Using
\begin{subequations}
    \begin{align}
        R_{\mu\nu\alpha\beta}S^{\nu\delta}S^{\sigma\mu} &= -\frac{1}{2}R_{\mu\nu\alpha\beta}S^{\sigma\delta}S^{\mu\nu}, \\
        \dot{p}_{\delta}S^{\nu\delta} &= -p_{\delta}\dot{S}^{\nu\delta},
    \end{align}
\end{subequations}
together with Eq.~\eqref{eq:ps}, we obtain
\begin{equation}
    m^2u^{\rho} = mp^{\rho} + \frac{2mR_{\mu\nu\alpha\beta} p^{\nu}S^{\rho\mu}S^{\alpha\beta}}{4m^{2}+R_{\mu\nu \alpha\beta}S^{\mu\nu}S^{\alpha\beta}}.
\end{equation}
Finally, the 4‑velocity can be expressed in terms of $p^{\mu}$ and $S^{\mu\nu}$~\cite{Tod:1976ud,Han:2016cdh} as:
\begin{equation}\label{eq:velocity}
     u^\alpha=-\frac{p_\lambda u^\lambda}{m^2}
     \left(p^\alpha+\frac{1}{2}
     \frac{S^{\alpha\beta}R_{\beta\gamma\mu\nu}p^\gamma S^{\mu\nu}}
     {m^2+R_{\mu\nu\rho\sigma}S^{\mu\nu}S^{\rho\sigma}/4}\right).
\end{equation}

For numerical integration, it is advantageous to replace the antisymmetric spin tensor $S^{\mu\nu}$ with a spin 4‑vector, following the procedure in Ref.~\cite{Singh:2005ha}:
\begin{equation}
    S^{\alpha} = \frac{1}{2m}{\varepsilon^{\alpha}}_{\beta\mu\nu}p^{\beta}S^{\mu\nu},
\end{equation}
where $\varepsilon_{\alpha\beta\mu\nu} = \sqrt{-g}\sigma_{\alpha\beta\mu\nu}$ is the Levi‑Civita tensor and $\sigma_{\alpha\beta\mu\nu}$ is the alternating symbol with $\sigma_{0123}=1$.
The inverse relation is
\begin{equation}
    S^{\alpha\beta} = \frac{1}{m}\varepsilon^{\alpha\beta\mu\nu}p_{\mu}S_{\nu},\label{eq:spin_tensor_from_vector}
\end{equation}
where $\epsilon^{\mu\nu\alpha\beta}=-\frac{1}{\sqrt{-g}}\sigma^{\mu\nu\alpha\beta}$. The magnitude of spin is conserved and can be written as
\begin{equation}\label{eq:ss}
    S^2 = S_{\alpha}S^{\alpha}=\frac{1}{2}S_{\mu\nu}S^{\mu\nu}.
\end{equation}
Then the TD-SSC~\eqref{eq:SSC} with a spin vector becomes
\begin{equation}\label{eq:newSSC}
    p_{\alpha}S^{\alpha} = 0,
\end{equation}
and the MPD equations with vector-form variables take a form ready for numerical integration:
\begin{subequations}\label{eq:neweom}
    \begin{align}
        \dot{x}^{\alpha} &=  -\frac{p_\lambda u^\lambda}{m^2}\left(p^\alpha+\frac{S^{\alpha\beta}R_{\beta\gamma\mu\nu}p^\gamma S^{\mu\nu}}{2\left(m^2+R_{\mu\nu\rho\sigma}S^{\mu\nu}S^{\rho\sigma}/4\right)}\right), \label{eq:eom_newx} \\
        \dot{p}^{\alpha} &= -{\Gamma^{\alpha}}_{\mu\nu} p^{\mu}u^{\nu} + \frac{1}{2m}{R^{\alpha}}_{\beta\rho\sigma}{\varepsilon^{\rho\sigma}}_{\mu\nu}S^{\mu}p^{\nu}u^{\beta}, \label{eq:eom_newp} \\
        \dot{S}^{\alpha} &= -{\Gamma^{\alpha}}_{\mu\nu} S^{\mu}u^{\nu} + \left( \frac{1}{2m^3}R_{\gamma\beta\rho\sigma}{\varepsilon^{\rho\sigma}}_{\mu\nu}S^{\mu}p^{\nu}S^{\gamma}u^{\beta} \right)p^{\alpha}. \label{eq:eom_news}
    \end{align}
\end{subequations}
Here, the spin tensor $S^{\alpha\beta}$ in Eq.~\eqref{eq:eom_newx} is understood as the expression~\eqref{eq:spin_tensor_from_vector}. Eqs.~\eqref{eq:neweom} together with the constraint~\eqref{eq:newSSC} constitute a closed system that we shall solve in the hairy BH background~\eqref{eq:metric}.

\subsection{Initial Conditions}\label{sub-1-3}

To numerically integrate the MPD equations, we must specify the initial conditions for the momentum $p^{\alpha}$ and the spin vector $S^{\alpha}$. Since the BH is much more massive than the test particle, the particle’s spin produces only a small correction to its geodesic motion, we therefore construct the initial 4‑momentum by first neglecting the spin, i.e., by solving the Hamilton-Jacobi equation for a spinless particle~\cite{kimpson_orbital_2020}:
\begin{equation}\label{eq:HJ}
    g^{\mu\nu} \frac{\partial W}{\partial x^{\mu}}\frac{\partial W}{\partial x^{\nu}} + m^2 = 0,
\end{equation}
where $W$ is the Jacobi action, $p_{\mu}\equiv \partial W/\partial x^{\mu}$ is the canonical momentum.

The spherical symmetry of the background spacetime admits two Killing vectors, $\partial_t$ and $\partial_\phi$, yielding the conserved energy $E=-p_t$ and the axial angular momentum $L_z=p_{\phi}$. A third constant of motion, the Carter constant $C$, is introduced to separate the radial and polar motions. The Jacobi action can therefore be written in the separated form
\begin{equation}
    W = -Et + \int{\frac{\sqrt{R(r)}}{r^{2}f(r)}\mathrm{d}r} + \int{\sqrt{\Theta(\theta)}\mathrm{d}\theta} + L_{z}\phi,
\end{equation}
with the functions $R(r)$ and $\Theta(\theta)$ given by
\begin{subequations}\label{eq:com}
    \begin{align}
        R(r) &= r^{4}E^{2} - r^{2}f(r)(r^{2}+L_{z}^{2}+C),\label{eq:R} \\
        \Theta(\theta) &= C-\frac{L_{z}^{2}}{\sin^2{\theta}}\cos^2{\theta}.\label{eq:Theta}
    \end{align}
\end{subequations}
From the Hamilton-Jacobi formalism, the 4-velocities ${x}^{\mu}$ can be expressed by
\begin{subequations} \label{eq:m-conditions}
    \begin{align}
        \dot{x}^0 &= \frac{E}{f(r)}, \\
        \dot{x}^1 &= \frac{\sqrt{R(r)}}{r^2}, \\
        \dot{x}^2 &= \frac{\sqrt{\Theta(\theta)}}{r^2}, \\
        \dot{x}^3 &= \frac{L}{r^{2}\sin^2{\theta}}.
    \end{align}
\end{subequations}
The initial 4‑momentum is then taken as
\begin{equation}
p^{\mu} = m\dot{x}^{\mu}.
\end{equation}
This geodesic initial condition is an approximation. Once the spin is turned on, the actual evolution obeys the full MPD equations. Meanwhile, the directions of the 4-momentum and the 4-velocity are not parallel. Especially, $E$, $L_z$ and $C$ will no longer be strictly conserved. The validity of this approximation stems from the fact that worldlines induced by different SSCs deviate quite slightly (much smaller than the scale of the orbit), allowing the dynamics to be dominated by the pole-dipole coupling rather than higher‑order dipole‑dipole terms. Consequently, the leading orders of 4‑velocity and 4‑momentum are parallel~\cite{kimpson_orbital_2020}.

Rather than specifying the initial data through a Keplerian orbital parametrization $(\mathcal{P}, e, \theta_{-})$, we follow Ref.~\cite{Giri:2022xkw} and prescribe the conserved quantities directly. Throughout this work we fix the angular momentum and the particle mass at $J=4$ and $m=1$, and choose the energy $E$ from the structure of the effective potential $V_{-}(r,\theta)$ derived in Sec.~\ref{sec-2}: the orbit is released from rest ($\dot{r}=0$) at the inner edge of the main bound band of $V_{-}(\cdot,\theta_{0})$ at the starting polar angle. For spinning particles we take $\theta_{0}=1.5$, slightly off the equatorial plane. For the spinless case, by contrast, the initial state is constructed analytically from the Hamilton-Jacobi solution above with $\theta_{0}=\pi/2$. The spin vector is chosen parallel to the total angular momentum, $\tilde{\theta}=0$, corresponding to the $V_{(-)}$ branch of the effective potential. Each trajectory is integrated for about 50 radial periods and sampled on $15000$ uniform-$\tau$ points; the sampling index $\tilde{t}$ is used as the time variable in all figures below.

To fully specify the initial conditions, the orientation of the spin vector must be prescribed. Following the approach of the laboratory frame for spinning test particles~\cite{kimpson_orbital_2020}, the spin is most conveniently parameterized in a locally 3-dimensional tetrad frame adapted to the spherical polar coordinates. In this local frame, the spin vector is decomposed as
\begin{subequations}\label{eq:local_S}
    \begin{align}
        \mathcal{S}^{\hat{1}} &= S\sin{\tilde{\theta}}\cos{\tilde{\phi}}, \\
        \mathcal{S}^{\hat{2}} &= -S\cos{\tilde{\theta}}, \\
        \mathcal{S}^{\hat{3}} &= S\sin{\tilde{\theta}}\sin{\tilde{\phi}}
    \end{align}
\end{subequations}
where $S$ denotes the constant spin magnitude, and the angles $\tilde{\theta}$, $\tilde{\phi}$ define its spatial orientation with respect to the local Cartesian axes. All local indices are written as superscripts, consistently with the transformation law below.

The spatial components of the spin vector in the global coordinate frame, $S^i$, are obtained by contracting the locally 3-dimensional tetrad components with the inverse tetrad vectors:
\begin{equation}\label{eq:tetrad transformation}
    S^i = {\mathcal{T}^{i}}_{\hat{\omega}}\mathcal{S}^{\hat{\omega}}.
\end{equation}
For a spherically symmetric tetrad, the high degree of symmetry allows the coordinate tetrad to be chosen in the simple diagonal form:
\begin{subequations}\label{eq:tetrads}
    \begin{align}
        {\mathcal{T}^{i}}_{\hat{1}} &= \bigl(0,\sqrt{g^{11}},0,0\bigr), \\
        {\mathcal{T}^{i}}_{\hat{2}} &= \bigl(0,0,\sqrt{g^{22}},0\bigr), \\
        {\mathcal{T}^{i}}_{\hat{3}} &= \bigl(0,0,0,\sqrt{g^{33}}\bigr). 
    \end{align}
\end{subequations}
Hence, the spatial components for the spin vector $S^{\mu}$ is obtained:
\begin{subequations}
    \begin{align}
        S^1 &= S\sin{\tilde{\theta}}\cos{\tilde{\phi}}\sqrt{f(r)}, \\
        S^2 &= -S\cos{\tilde{\theta}}/r, \\
        S^3 &= S\sin{\tilde{\theta}}\sin{\tilde{\phi}}/(r\sin{\theta}), 
    \end{align}
\end{subequations}
while the time component should be determined by the TD-SSC (Eq.~\ref{eq:SSC}):
\begin{subequations}\label{eq:S0}
    \begin{align}
        S^0 &= -(p_{1}S^{1}+p_{2}S^{2}+p_{3}S^{3})/p_0.
    \end{align}
\end{subequations}

\subsection{Orbital Dynamics}\label{sub-orbitalD}\label{sub-1-4}

We now integrate the MPD equations for hair parameters $h=0$, $0.1$, $0.2$ and $0.3$, using the initial-data prescription of subsection~\ref{sub-1-3}: $J=4$, $m=1$, with the energy fixed by the effective-potential structure and the orbit released from rest at the inner edge of the bound band. This range of $h$ lies below the critical value $h_{\rm crit}(S)$ at every spin considered here, so that a bound region still exists; see Sec.~\ref{sub-eff-p}.

\begin{figure}[htbp]
\includegraphics[scale=0.45]{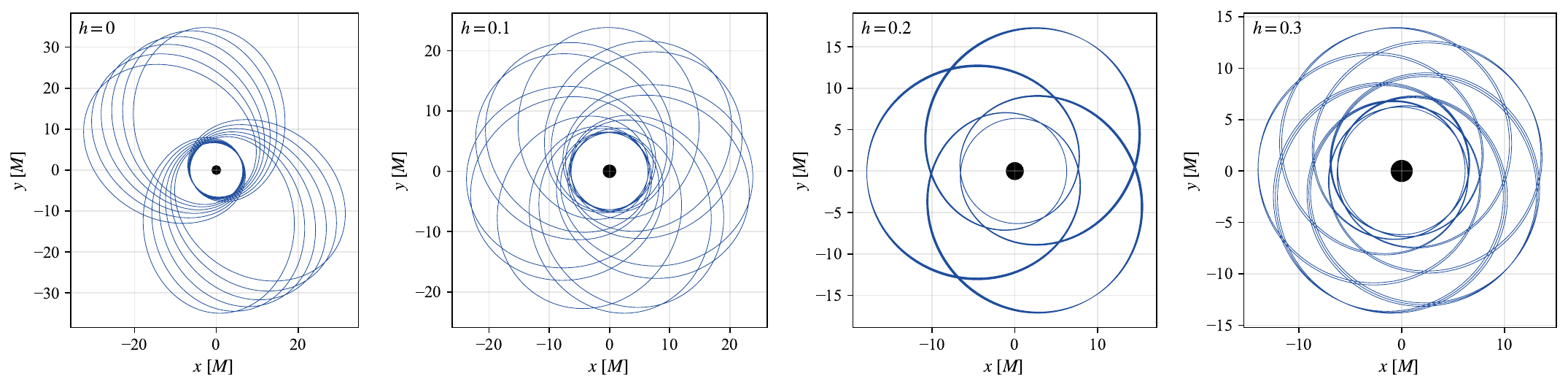}\centering
\caption{Orbital evolution of a spinless particle ($S=0$) in the equatorial plane for $h=0$, $0.1$, $0.2$ and $0.3$. The filled black disk indicates the horizon region (drawn at half the horizon radius for clarity). Only the first 12 radial periods of each trajectory are shown.}
\label{fig:orbits(S=0)}
\end{figure}

As a baseline we begin with the spinless particle ($S=0$). Fig.~\ref{fig:orbits(S=0)} shows the resulting equatorial orbits, each trajectory is a precessing rosette whose pericentre advances from cycle to cycle, filling an annulus bounded by the pericentre and apocentre of the bound band. As $h$ increases the precession becomes more rapid, and the rosette is wound more tightly—the same frequency growth that will reappear in the waveform analysis of Sec.~\ref{sec-4}. The motion remains strictly planar, as required by the spherical symmetry of the background.

\begin{figure}[htbp]
\centering
\includegraphics[scale=0.575]{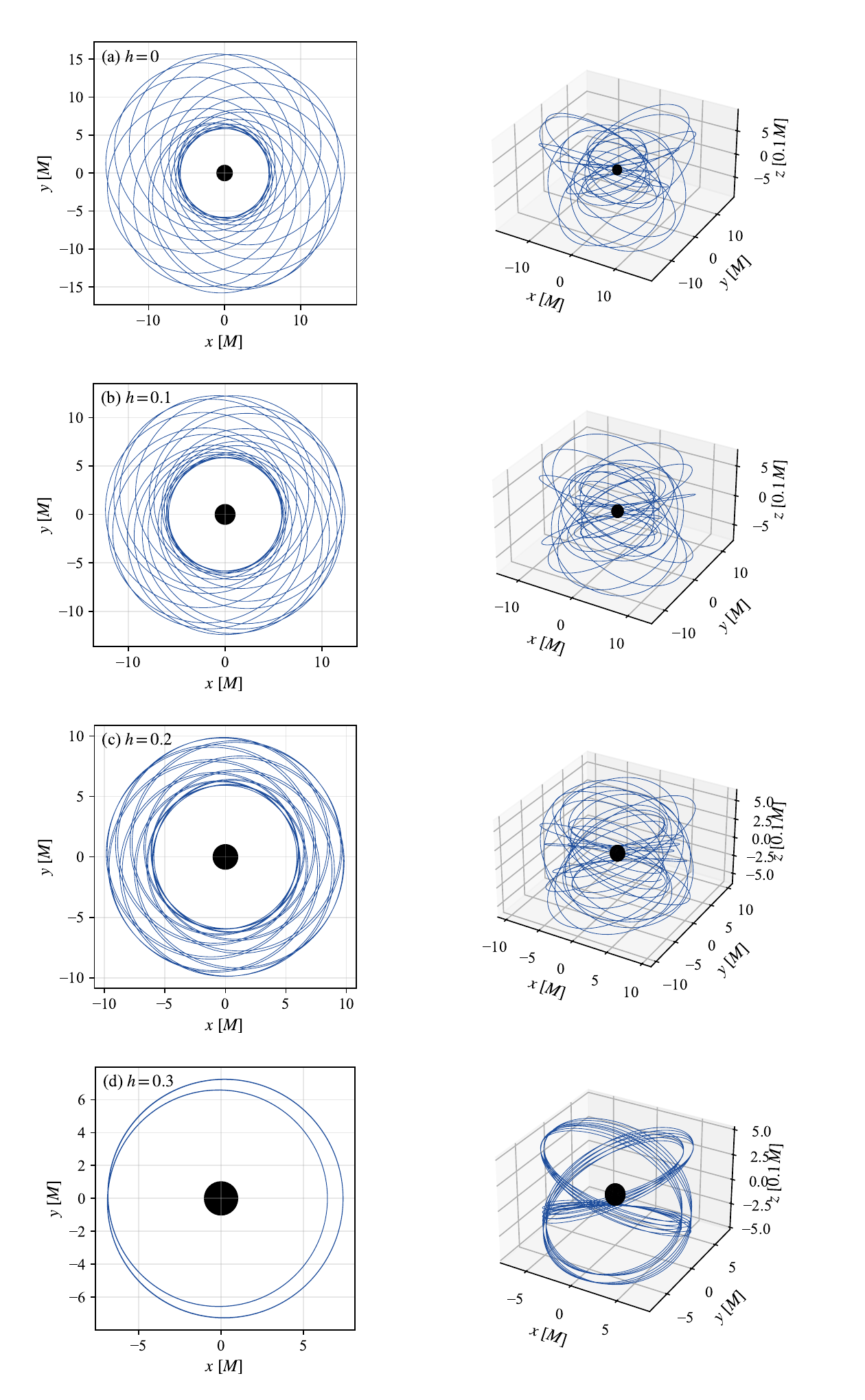}
\caption{Orbital evolution of a spinning particle ($S=0.5$) for $h=0$, $0.1$, $0.2$ and $0.3$ (top to bottom). Left column: $x$--$y$ projections. Right column: full three-dimensional trajectories, drawn with an expanded vertical scale to make the off-plane structure visible. Only the first 12 radial periods are shown.}
\label{fig:orbits(S=0.5)}
\end{figure}

With spin turned on ($S=0.5$, Fig.~\ref{fig:orbits(S=0.5)}), the particle lifts off the equatorial plane: the three-dimensional trajectories oscillate about the equator with a small but finite vertical extent, a direct consequence of the spin-curvature coupling in the MPD equations. In the $x$--$y$ projection the orbits also develop a petal-like precession pattern, originating from the periodic variation of the maximal radial distance between successive precession cycles. As $h$ increases, the bound band of the effective potential shrinks (cf.\ Sec.~\ref{sub-eff-p}), the radial annulus narrows, and the petal structure gradually fades.

\begin{figure}[htbp]
\centering
\includegraphics[scale=0.575]{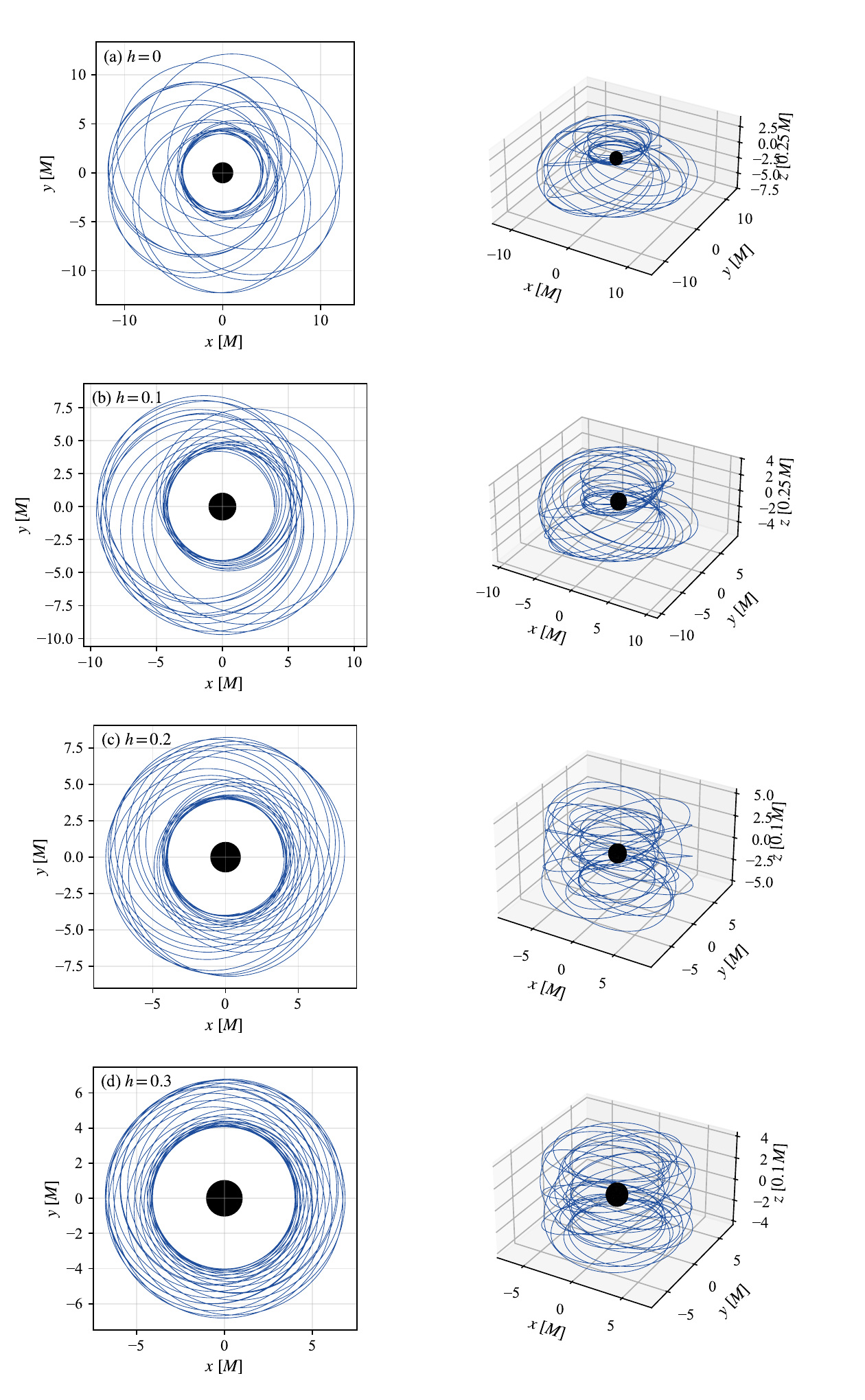}
\caption{Orbital evolution of a spinning particle ($S=1$) for $h=0$, $0.1$, $0.2$ and $0.3$ (top to bottom). Left column: $x$--$y$ projections. Right column: full three-dimensional trajectories, drawn with an expanded vertical scale to make the off-plane structure visible. Only the first 12 radial periods are shown.}
\label{fig:orbits(S=1)}
\end{figure}

Increasing the spin to $S=1$ (Fig.~\ref{fig:orbits(S=1)}) enlarges the vertical excursion further and renders the passages between successive pericentres visibly irregular. The trajectory now samples the neighbourhood of the off-plane saddle points of the type-(B2) potential---the seat of the chaotic dynamics analysed in Sec.~\ref{sec-3}. The contraction of the accessible band with increasing $h$ is common to all spins; what distinguishes the chaotic response to the hair is not visible in the orbital morphology alone and requires the dynamical indicators of Sec.~\ref{sec-3}.

To quantify the deviation from the equatorial plane, we use the measure
\begin{equation}
   \tilde{r} = r - r\sin{\theta},
\end{equation}
where $r\sin{\theta}$ is the projected radius in the $x$--$y$ plane, so that $\tilde{r}=0$ indicates confinement to the equatorial plane. Note that the variable $\tilde{t}$ used here and hereafter does not represent the true dynamical time of the system, but merely denotes the sampling points.

\begin{figure}[htbp]
\centering
\includegraphics[scale=0.55]{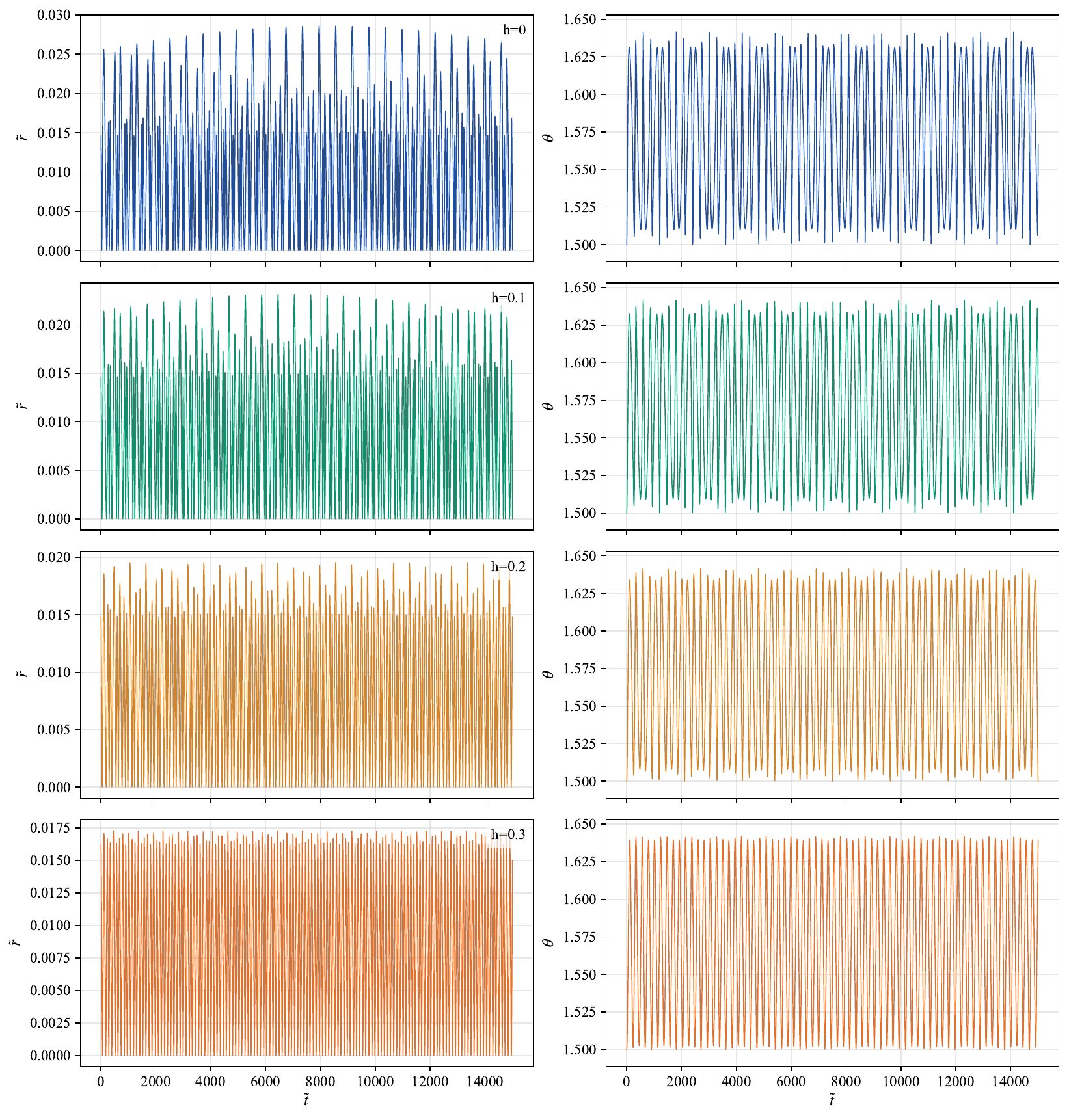}
\caption{$\tilde{r}$ (left column) and the polar angle $\theta$ (right column) as functions of the sampling index $\tilde{t}$, for $S=0.5$ and $h=0$, $0.1$, $0.2$, $0.3$ (top to bottom). The $\theta$ axis spans the band actually visited by the orbit.}
\label{fig:periodical_variables}
\end{figure}

\begin{figure}[htbp]
\centering
\includegraphics[scale=0.55]{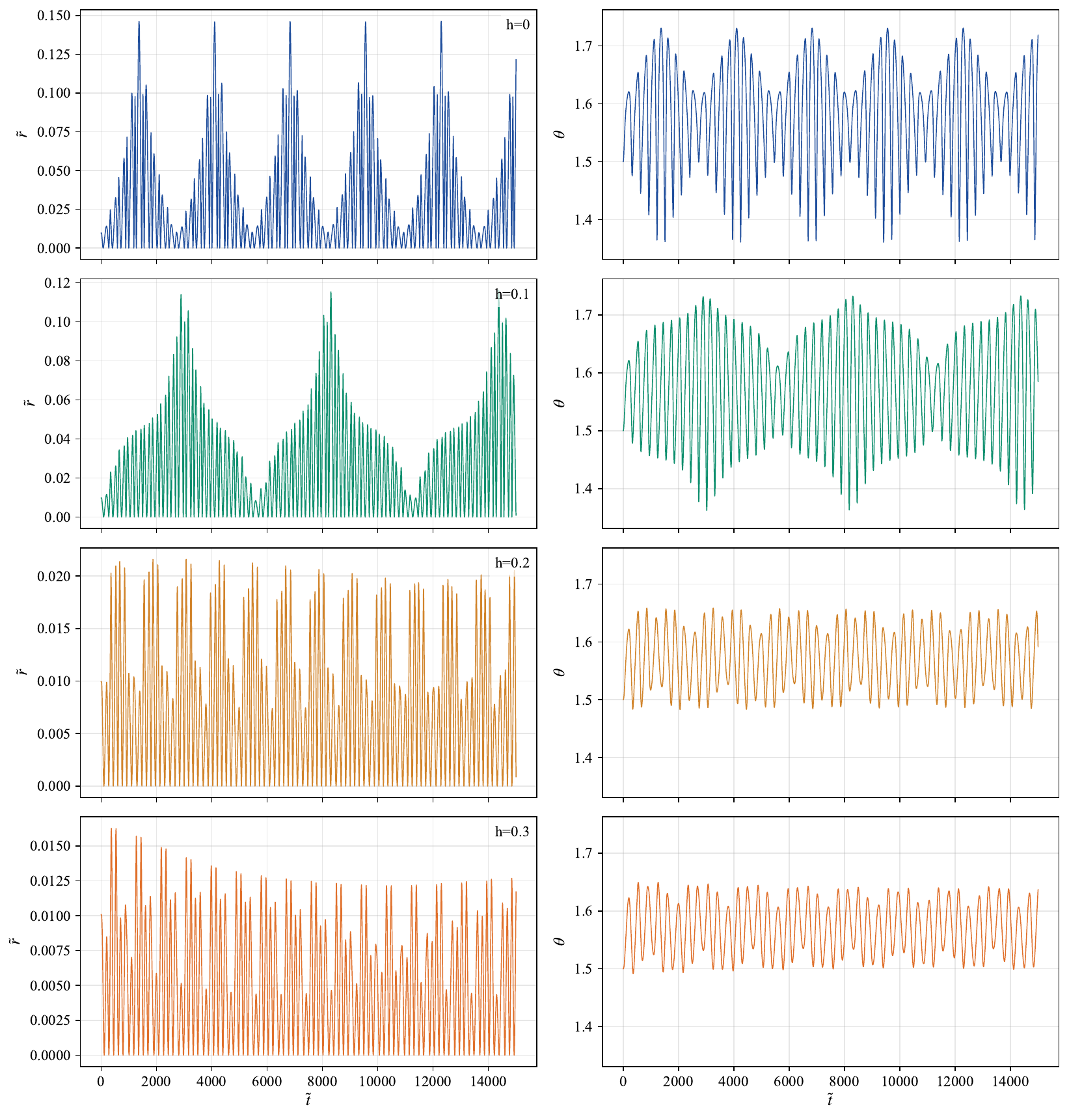}
\caption{$\tilde{r}$ (left column) and the polar angle $\theta$ (right column) as functions of the sampling index $\tilde{t}$, for $S=1$ and $h=0$, $0.1$, $0.2$, $0.3$ (top to bottom). The $\theta$ axis spans the band actually visited by the orbit.}
\label{fig:periodical_variables_S1}
\end{figure}

Figs.~\ref{fig:periodical_variables} and \ref{fig:periodical_variables_S1} display the evolution of $\tilde{r}$ and $\theta$ for $S=0.5$ and $S=1$. For $S=0.5$ both $\tilde{r}$ and $\theta$ oscillate quasi-periodically, with amplitudes that decrease and frequencies that grow with $h$—the orbit is pressed onto an increasingly tight, regular band. For $S=1$ the picture changes qualitatively: the $\theta$ oscillations remain confined to a narrow band around the equator (roughly $1.35\lesssim\theta\lesssim1.78$), but the $\tilde{r}$ signal develops strong, irregular bursts concentrated around the pericentre passages, where the orbit grazes the saddle region of the potential. These bursts are the morphological signature of the radial chaos that the LEs and Poincar\'e sections of Sec.~\ref{sec-3} will quantify.

\section{Effective Potential Analysis}\label{sec-2}

Before presenting the explicit form of the effective potential, we first clarify the sense in which this term is used for a spinning test particle. Owing to the spin–orbit coupling in the equations of motion~\eqref{eq:eom_newp}, the standard effective potential cannot be defined in the usual way~\cite{Suzuki:1996gm}. Nevertheless, following the terminology of Ref.~\cite{Suzuki:1996gm}, we shall define the boundary contour that delimits the allowed region of motion by imposing the condition that both the radial and the polar momentum vanish. This contour of zero meridional momentum serves as an ``effective potential'' for the spinning particle.

To illustrate the explicit effect of spin, we now construct this effective potential for a spinning test particle in the Horndeski hairy BH spacetime.

\subsection{Effective Potential}

Our first step is to derive analytic expressions for each component of the spin tensor. This task is tractable because of the spacetime’s spherical symmetry. For a spinning particle, a conserved quantity $Y$ along its worldline can be constructed from the Killing vector $\xi^{\mu}$ of the background~\cite{Suzuki:1996gm}. It is defined as
\begin{equation}
    Y=\xi^{\mu}p_{\mu}-\frac{1}{2}\xi_{\mu;\nu}S^{\mu\nu}, \label{eq:ConsQ}
\end{equation}
where the semicolon denotes covariant differentiation.

Using the Killing vectors associated with the static and axial symmetries, $\xi^{\mu}_{(t)}=(1,0,0,0)$ and $\xi^{\mu}_{(\phi)}=(0,0,0,1)$, we obtain two constants of motion:
\begin{subequations}
    \begin{align}
        E &= -Y_{(0)} = -p_0 - \frac{1}{2}f'(r)S^{01},\label{EY} \\
        L_z &= Y_{(3)} = p_3 + r(rS^{23}\cos{\theta}\sin{\theta}+S^{13}\sin^2{\theta}).
    \end{align}
\end{subequations}
The spherical symmetry also guarantees conservation of the remaining Cartesian components of the angular momentum. The corresponding Killing vectors are given by
\begin{equation}
  \xi^{\mu}_{(x)}=(0,0,-\sin{\phi},-\cot{\theta}\cos{\phi}),\,\,\,\,\xi^{\mu}_{(y)}=(0,0,\cos{\phi},-\cot{\theta}\sin{\phi}). \label{eq:KillingLxy}
\end{equation}
Inserting Eq.~\eqref{eq:KillingLxy} into Eq.~\eqref{eq:ConsQ}, we have
\begin{subequations}
    \begin{align}
         L_x={}&-p_2\sin\phi-p_3\cot\theta\cos\phi-rS^{12}\sin\phi\notag\\
         &-rS^{13}\sin\theta\cos\theta\cos\phi
         +r^2S^{23}\sin^2\theta\cos\phi,\\
         L_y={}&p_2\cos\phi-p_3\cot\theta\sin\phi+rS^{12}\cos\phi\notag\\
         &-rS^{13}\sin\theta\cos\theta\sin\phi
         +r^2S^{23}\sin^2\theta\sin\phi..
    \end{align}
\end{subequations}
Without loss of generality we align the total angular momentum with the $z$-axis, i.e. we set $(L_{x}, L_{y}, L_{z})=(0, 0, L)$. This choice yields the following solutions for the spatial components of the spin tensor:
\begin{subequations}
    \begin{align}
        S^{12} &= -\frac{p_2}{r}, \\
        S^{13} &= \frac{1}{r}\left( L-\frac{p_3}{\sin^2{\theta}} \right), \\
        S^{23} &= \frac{L\cot{\theta}}{r^2}.
    \end{align}
\end{subequations}
The remaining components $S^{0i}$ are determined from the TD-SSC~\eqref{eq:SSC}. Substituting the above results gives
\begin{subequations}
    \begin{align}
        S^{01} &= -\frac{1}{rp_0}\left[ (p_2)^2 + \frac{(p_3)^2}{\sin^2{\theta}} - Lp_3 \right], \\
        S^{02} &= \frac{1}{rp_0}\left( p_{1}p_{2} + \frac{Lp_3}{r}\cot{\theta} \right), \\
        S^{03} &= -\frac{1}{rp_0}\left( Lp_1 - \frac{p_{1}p_{3}}{\sin^2{\theta}} + \frac{Lp_2}{r}\cot{\theta} \right).
    \end{align}
\end{subequations}
Inserting the expression for $S^{01}$ into the conserved energy equation~\eqref{EY} leads to a compact formula for the energy:
\begin{equation}\label{eq:E}
    E = -p_0 + \frac{(p_{2})^2 - Lp_3 + (p_{3})^2\csc^2{\theta}}{2p_{0}r}f'(r).
\end{equation}

We now implement the definition of the effective potential. Owing to the spin–orbit coupling in the
equations of motion, the standard effective potential cannot be defined in the usual. Thus, following Ref.~\cite{Suzuki:1996gm}, the boundary contour of zero meridional momentum can serve as an “effective potential”. We first restrict the spin to lie in the meridional plane, which corresponds to setting $p^1 = 0$ and $p^2 = 0$ at the turning points of the motion. Under this assumption we parameterize the remaining momentum components as
\begin{subequations}\label{eq:p0p3}
    \begin{align}
        p_0 &= -mf^{1/2}(r)\cosh{X(r,\theta)}, \\
        p_3 &= mr\sin{\theta}\sinh{X(r,\theta)},
    \end{align}
\end{subequations}
where $X(r,\theta)$ is a function to be determined. Substituting Eqs.~\eqref{eq:p0p3} into the spin tensor solutions obtained above yields the following forms
\begin{subequations}\label{eq:sol_S}
    \begin{align}
        S^{01} &= S^{01}=\frac{\tanh X(r,\theta)}{\sqrt f(r)}[mr\sinh X(r,\theta)-L\sin\theta], \\
        S^{02} &= -\frac{L\cos{\theta}}{r\sqrt{f(r)}}\tanh{X(r,\theta)}, \\
        S^{03} &= 0,\\
        S^{12} &= 0, \\
        S^{13} &= \frac{L}{r} - m\csc{\theta}\sinh{X(r,\theta)}, \\
        S^{23} &= \frac{L\cot{\theta}}{r^2}.
    \end{align}
\end{subequations}

To determine $\sinh{X(r,\theta)}$, we first insert the given spin tensor into Eq.~\eqref{eq:ss}, yielding the expanded form:
\begin{equation}
    S^2 = \frac{1}{2}S_{\mu\nu}S^{\mu\nu} = -{(S_{01})}^{2} + \frac{r^{2}{(S_{12})}^{2}}{f(r)} - r^{2}{(S_{02})}^{2}f(r) + r^{4}{(S_{23})}^{2}\sin^{2}\theta + \frac{r^{2}{(S_{13})}^{2}\sin^{2}\theta}{f(r)} - r^{2}{(S_{03})}^{2}f(r)\sin^{2}\theta.
\end{equation}
Using Eqs.~\eqref{eq:sol_S}, the conserved spin relation can be recast as:
\begin{equation}\label{eq:conservedspin}
    -S^{2}f(r) + \left[ L^{2} \cos^2{\theta}f(r) + (L\sin{\theta}-mr\sinh{X(r,\theta)})^2 \right]/\cosh^2{X(r,\theta)} = 0 ,
\end{equation}
and $\sinh{X(r,\theta)}$ is then directly obtained:
\begin{subequations}
    \begin{align}
        \sinh{X(r,\theta)} &= \frac{\omega(r,\theta) \pm \sqrt{\omega^2(r,\theta)-\varsigma(r)[(L^2-S^2)f(r)+L^{2}\sin^2{\theta}(1-f(r))]}}{\varsigma(r)}, \\
        \omega(r,\theta) &= mrL\sin{\theta}, \\
        \varsigma(r) &= m^{2}r^{2} - S^{2}f(r).
    \end{align}
\end{subequations}

Finally, inserting $\sinh X(r,\theta)$ into the energy expression~\eqref{eq:E} gives the effective potential $V$ defined by the condition $p^1 = p^2 = 0$:
\begin{subequations}
    \begin{align}
        E &= V_{(\pm)}(r,\theta,L,S) = m\sqrt{f(r)}\sqrt{1+y_\pm^2}+\frac{f'(r)(L\sin\theta-mry_\pm)y_\pm}{2\sqrt{f(r)}\sqrt{1+y_\pm^2}}\\
        y &= \sinh{X(r,\theta)}.
    \end{align}
\end{subequations}
The subscript $(\pm)$ corresponds to the orientation of the particle’s spin relative to the total angular momentum $L$: $V_{(-)}$ describes the configuration where the spin is parallel to $L$, while $V_{(+)}$ corresponds to the antiparallel case. In what follows we adopt the parallel configuration $V_{(-)}$, since it allows the particle to approach the horizon more closely and is therefore more likely to exhibit strong spin effects~\cite{Suzuki:1996gm}.

Following the classification of effective potentials in Ref.~\cite{Suzuki:1996gm}, there are four distinct types: restrictive (B1, B2) and non-restrictive types (U1, U2). Type (B1), characterized by large angular momentum $L$ and relatively small spin $S$, exhibits no chaos. As the spin increases, the system switches to type (B2), where chaotic behavior emerges. Both non-restrictive types are associated with small $L$: type (U1) involves low spin, while type (U2) features higher spin; neither exhibits chaos. The four types and their key features are summarized in Table~\ref{tables-eff-p}. 

Since chaotic motion is associated with the high‑spin restrictive type (B2), we primarily focus on the case $S=1$, together with the strongly chaotic extension to $S=1.5$ (beyond the commonly used pole-dipole range), which are expected to display strong chaotic behavior. In the following Sec.~\ref{sec-3}, we investigate how the Horndeski hair influences this chaotic dynamics.

\begin{table}[h]
  \centering
  \label{tab:potential_class}
  {
  \begin{tabular}{|c|c|c|c|c|}
    \hline
   \ Category \ & \ Potential Type \ & \ Angular Momentum ($L$) \ & Spin ($S$) & \ Key Features \ \\
    \hline
    B1 & Restrictive & Large & Low & No chaos \\
    B2 & Restrictive & Large & Higher ($>$\,B1) & chaos   \\
    U1 & Non-restrictive & Small & Low & No chaos   \\
    U2 & Non-restrictive & Small & Higher ($>$\,U1) & No chaos   \\
    \hline
  \end{tabular}%
  }
  \caption{Classification of the effective potentials.}
  \label{tables-eff-p}
\end{table}

\subsection{Effective Potential Metaphor and Saddle Points}\label{sub-eff-p}

\begin{figure}[htbp]
	\subfigure[\ $S=0$]{
		\begin{minipage}{1\linewidth}
            \includegraphics[scale=0.275,bb=75 0 550 400]{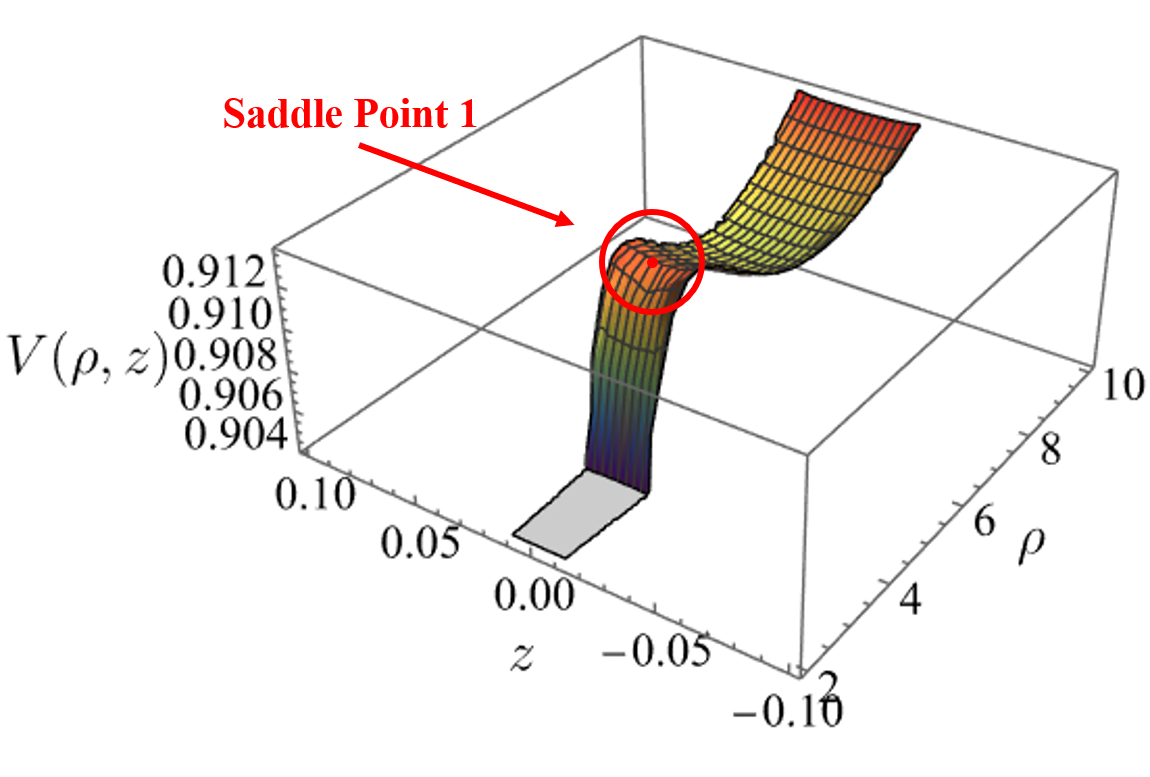}
            \includegraphics[scale=0.275,bb=0 0 550 400]{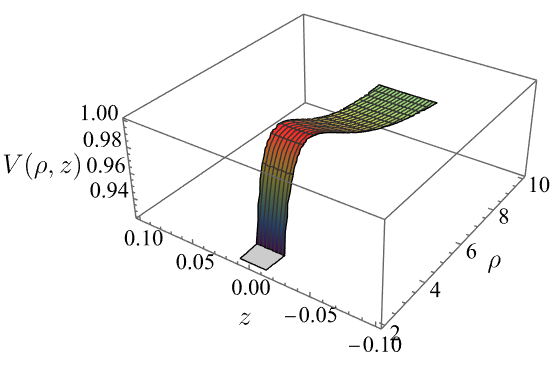}
            \includegraphics[scale=0.275,bb=0 0 550 400]{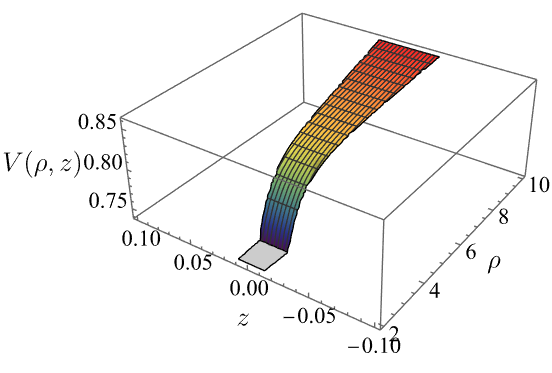}
		\end{minipage}
		}
    \subfigure[\ $S=0.5$]{
		\begin{minipage}{1\linewidth}
            \includegraphics[scale=0.23,bb=400 0 250 400]{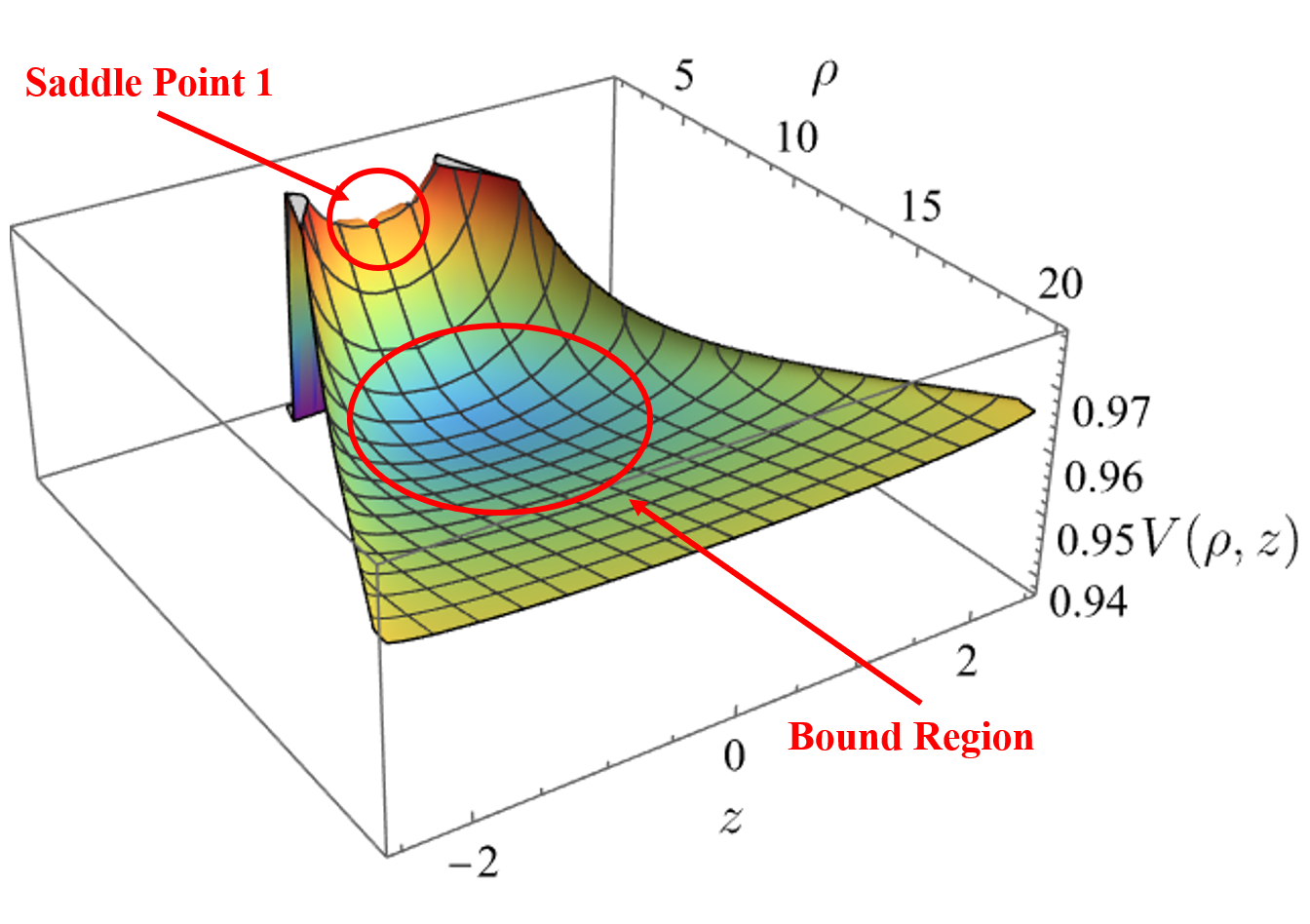}
            \includegraphics[scale=0.23,bb=-400 0 250 400]{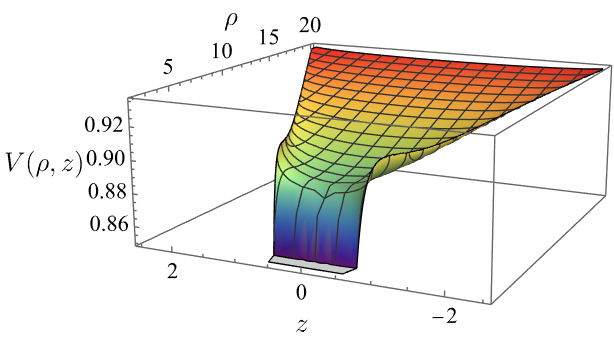}
            \includegraphics[scale=0.23,bb=-400 0 250 400]{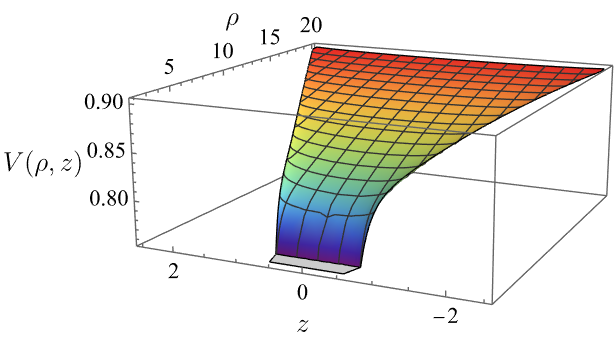}
		\end{minipage}
	}
    \subfigure[\ $S=1$]{
		\begin{minipage}{1\linewidth}
            \includegraphics[scale=0.23,bb=400 0 250 400]{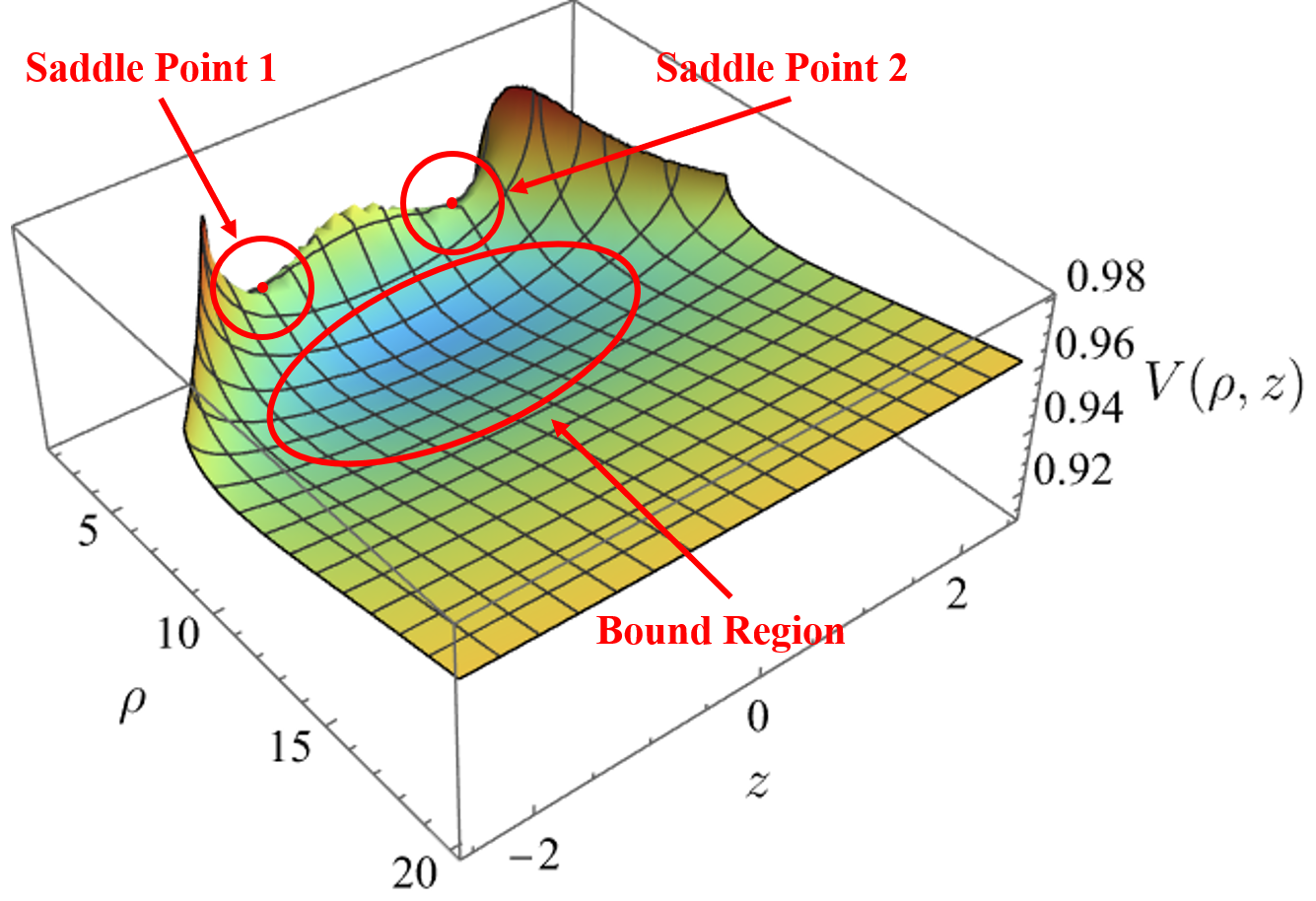}
            \includegraphics[scale=0.23,bb=-400 0 250 400]{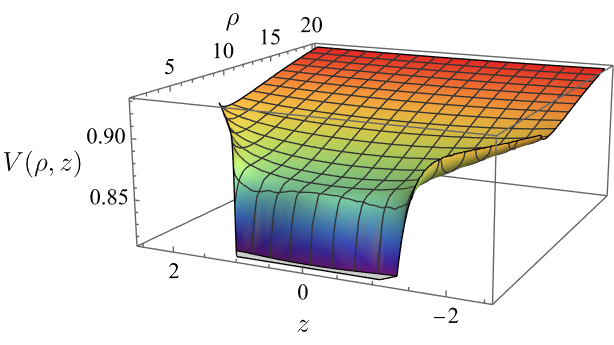}
            \includegraphics[scale=0.23,bb=-400 0 250 400]{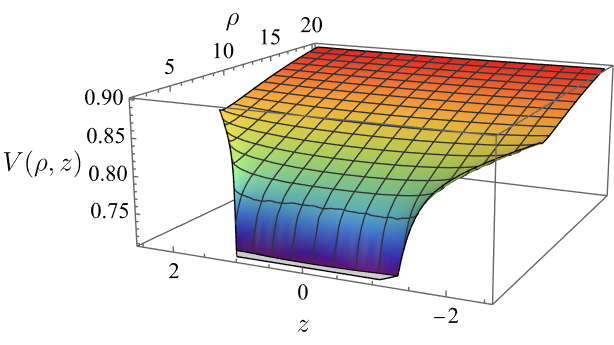}
		\end{minipage}
	}
	\caption{Effective potentials in cylindrical coordinates, varying over the hairy parameter $h$ (columns: $0,0.5,1$) and the particle spin $S$ (rows: $0,0.5,1$).
    }\label{fig:V_cyl}
\end{figure}

In this subsection, we analyze the effective potential in cylindrical coordinates $(\rho, z)$ (see Fig.~\ref{fig:V_cyl}) to elucidate the orbital dynamics from the preceding subsection and the ensuing chaotic behavior. Our discussion is organized by the particle spin $S$, with each value ($0$, $0.5$, $1$) illustrating a key stage in the transition to chaos, and examines the impact of the hairy parameter $h$ ($0$, $0.5$, $1$). The key properties of the potential are summarized below.
\begin{itemize}
\item \textbf{No scalar hair ($h=0$).}
\begin{itemize}
    \item \textbf{Spinless case ($S=0$).} The potential is sharply confined near $z=0$ and attains its extremum on the equatorial plane ($z=0$), forming a single saddle point that confines the particle motion. This structure, characteristic of a type (B1) potential, explains the planar orbits observed in Fig.~\ref{fig:orbits(S=0)}.
    \item \textbf{Low-spin case ($S=0.5$).} With spin, the potential extends its distribution and develops a barrier along the $z$-direction near the horizon, permitting off-plane motion. While a single saddle point on the equatorial plane maintains a bound region (still type (B1)), the $z$-barrier allows the particle to reach varying maximal distances in successive precession cycles by “sliding” from the barrier, leading to the petal-like $x$–$y$ projection.
    \item \textbf{Maximal-spin case ($S=1$).} At maximal spin $S=1$, the equatorial saddle point bifurcates into two, located symmetrically off the plane. This creates a new bound region, characteristic of a type (B2) potential, within which the particle executes complex retrograde motion — the direct source of the observed chaos.
    \end{itemize}
\item \textbf{Effect of the hairy parameter.} Increasing $h$ flattens the potential, lowering barriers, erasing saddle points, and ultimately destroying the bound regions. This reduces the variation in maximal precession distance, explaining the fading of the petal-like orbits in Figs.~\ref{fig:orbits(S=0.5)} and \ref{fig:orbits(S=1)}. The system is notably sensitive for $S=1$: the off-plane saddle points coalesce with the degenerate equatorial (barrier, well) pair at $h_{\text{crit}} = 0.4927$ for $J=4$. For larger $h$, the potential becomes monotonic, resembling a non-restrictive type (U).
\end{itemize}

\begin{figure}[htbp]
\centering
\includegraphics[scale=0.625]{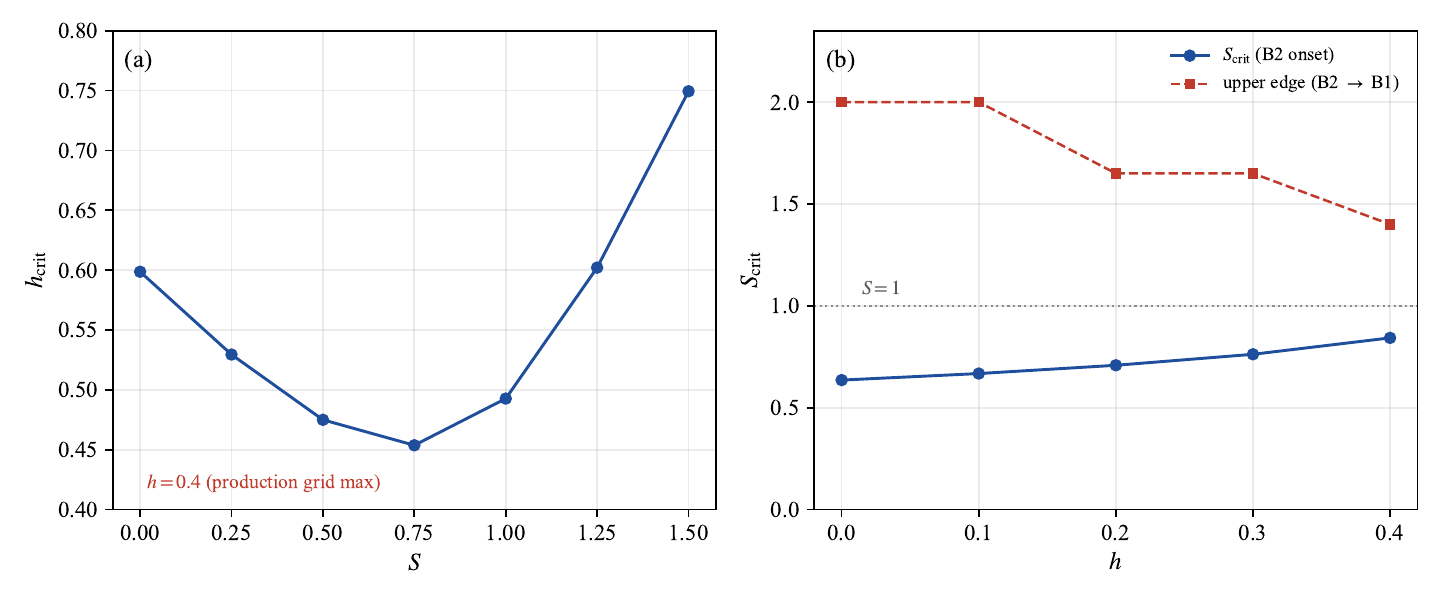}
\caption{Critical boundaries of the (spin, hair) plane at $J=4$. (a)~$h_{\text{crit}}(S)$: the equatorial barrier/well degeneracy of $V_{-}$, beyond which the potential is monotonic and no bound orbit exists. (b)~$S_{\text{crit}}(h)$: the smallest spin at which the off-plane saddle pair appears (the B1$\to$B2 transition), together with the large-spin upper edge where the B2 structure closes. The dashed line in (a) marks $h=0.4$, the largest hair value used in this work; the dotted line in (b) marks the primary spin $S=1$.}
\label{fig:hcrit}
\end{figure}

The value of $h_{\text{crit}}$ marks the practical boundary of the present analysis: beyond it the equatorial profile $V_{-}(r,\pi/2)$ is monotonic, no pair of turning points exists at any energy, and no bound orbit—with physical meaning—can be constructed in the ($E$, $J$)-prescribed setup. Importantly, $h_{\text{crit}}$ is not a universal number but depends on the spin. Fig.~\ref{fig:hcrit}(a) shows $h_{\text{crit}}(S)$ obtained by bisecting on the barrier-well gap at $J=4$: the curve is U-shaped, decreasing from $h_{\text{crit}}=0.599$ at $S=0$ to a minimum of $0.454$ near $S\simeq0.75$, and rising again to $0.493$ at $S=1$ and $0.749$ at $S=1.5$. Moderate spin thus makes the barrier-well structure most fragile against the hair, while strong spin re-stiffens it. A further nuance appears at large spin: for $S=1.5$ the off-plane saddle points dive into the equatorial plane already at $h\simeq0.36$—rendering the equatorial well unstable against polar escape—while the one-dimensional equatorial pair survives up to $h_{\text{crit}}=0.749$. All hair values used in this work lie below the corresponding $h_{\text{crit}}(S)$ of their spin.

The companion boundary---the smallest spin $S_{\text{crit}}$ at which the off-plane saddle pair appears at fixed $h$ (the B1$\to$B2 transition)---is equally hair-dependent [Fig.~\ref{fig:hcrit}(b)]: $S_{\text{crit}}$ rises from $0.636$ at $h=0$ to $0.843$ at $h=0.4$, because the hair erodes the saddle structure that the spin must create. The chaos-capable B2 region is a finite band in spin: at very large spin the spin--orbit term re-flattens the potential and the band closes (the upper edge falls from $S\simeq2$ at $h=0$ to $1.4$ at $h=0.4$, verified with full-resolution scans of $V_{-}$). The production spins sit on well-defined sides of this boundary: $S=0.5$ lies below $S_{\text{crit}}$ at every $h$ (type B1, explaining its small Lyapunov exponents in Sec.~\ref{sec-3}), $S=1$ lies inside the band at every $h$, and $S=1.5$ exits the band exactly at $h=0.4$ (its saddles dive into the equator, as noted above).

\section{Chaotic Dynamics Analysis}\label{sec-3}

A key signature of chaotic dynamics is a positive LE $\lambda$, which quantifies the average rate of divergence between nearby trajectories. When the numerically computed maximal LE converges to a positive value, it reliably indicates the presence of chaos in the system. Meanwhile, in the procedure of calculating the LE, we will rebuild the Lorenz attractor, which is a set encompassing all possibly dynamical states of a particle in long-time evolution. 

Another crucial indicator of chaos is the Poincaré section, a lower-dimensional projection of the phase space. Chaotic motion is characterized by a scatter of discrete points on the Poincaré section, in contrast to the smooth, closed curve produced by regular motion. This occurs because the trajectory, due to its chaotic nature, never returns exactly to a previous point after one period, and each intersection with the section lands at a distinct location. Over sufficient time, these points can form intricate fractal structures.

In this section, we investigate how the scalar hair parameter $h$ influences these chaotic indicators for three spin values, $S=0.5$, $1$ and $1.5$. The case $S=1.5$ lies beyond the commonly used range $S\lesssim1$ of the pole-dipole approximation; we include it as a controlled probe of the strong spin-curvature coupling regime, where the effects of the hair on the chaos can be examined in their most pronounced form.
To this end, we first reconstruct the phase space of the system via Takens’ embedding theory in subsection \ref{sub-TET}. We then introduce the Wolf algorithm for computing LEs in subsection \ref{sub-WALE}. In subsection \ref{sub-ICD}, we numerically calculate these chaotic indicators and analyze the chaotic properties of the system.

\subsection{Takens' Embedding Theory}\label{sub-TET}

Having obtained the numerical data by integrating the MPD equations, we can reconstruct the phase space of the system using Takens’ embedding theorem~\cite{botvinick2024measure}. According to Takens' theory, a time series of a single dynamical variable contains sufficient information to reconstruct a new phase space that is diffeomorphic to the original one~\cite{sano1985measurement}. Let the true (unknown) state of the dynamical system be represented by a vector in the original phase space
\begin{equation}
\mathbf{A}(t) = \bigl(\alpha_{1}(t),\alpha_{2}(t),\dots ,\alpha_{\mathcal{D}}(t)\bigr),
\end{equation}
where $\mathcal{D}$ is the dimension of the original phase space. Suppose we measure a scalar observable
\begin{equation}
    s(t) = \mathcal{M}(\alpha_{1}(t)),
\end{equation}
where $\mathcal{M}$ is a smooth measurement function. According to Takens’ embedding theorem, a topologically equivalent phase space can be reconstructed from the time series $s(t)$ via delay coordinates
\begin{equation}\label{eq:new PS}
\mathbf{B}(t) = \bigl(s(t),s(t+T),s(t+2T),\dots ,s(t+(c-1)T)\bigr).
\end{equation}
Here, $T$ is the time delay, chosen as an integer multiple of the sampling interval $\Delta t$. The embedding dimension $c$ must satisfy $c>2D$, $D$ is the dimension of the original Lorenz attractor. The mapping $\Phi: \mathbf{A}(t)\to \mathbf{B}(t)$ is then an embedding. $\Phi$ and its inverse $\Phi^{-1}$ are smooth one-on-one mappings and diffeomorphisms, preserving invariant quantities such as Lorenz attractor and LEs.

A useful analogy is viewing a three‑dimensional object under varying illumination: its one‑dimensional shadow changes with the light direction. Takens’ theorem guarantees that if we record the shadow from sufficiently many independent viewpoints, namely an adequate embedding dimension $c$, the full object can be uniquely reconstructed. In practice, one should choose an observable that best reflects the irregular, chaotic nature of the dynamics. Here we use the radial coordinate $r(t)$ as the measured variable. Other scalar functions that depend nonlinearly on the state, e.g., kinetic or potential energy, could also serve as suitable observables.

\subsection{Wolf Algorithm for LEs}\label{sub-WALE}

To compute the LEs numerically we employ the Wolf algorithm~\cite{wolf1985determining}. Based on the new phase space~\eqref{eq:new PS}, the Wolf algorithm can numerically implement the definition of LE. Its main idea is to trace a long-time evolution of elements in an infinite small bulk. Our goal is to approach the maximal LE:
\begin{equation}
    \lambda_{max} = \lim_{t\to \infty} \lim_{|\delta\mathbf{Z}(0)|\to0}\frac{1}{t}\log_{2}\frac{|\delta\mathbf{Z}(t)|}{|\delta\mathbf{Z}(0)|},
\end{equation}
where $\delta\mathbf{Z}(t)$ is the deviate vector between two points in the reconstructed phase space at time $t$. 

The procedure begins by choosing an initial point $\Omega(t_0)$ from Eq.~\eqref{eq:new PS} on a fiducial trajectory. Within a small sphere of radius $d$ centered at this point, we then locate its closest neighbour $\Omega({t_0}')$, which lies on a distinct trajectory corresponding to a different state of the system. The neighbor is identified as the point that minimizes the Euclidean distance in the reconstructed phase space, yielding an initial separation $L_0 = ||\Omega(t_0)-\Omega({t_0}')|| < d$. The appropriate value of the search radius $d$ depends on the purity of the data: cleaner data (high signal‑to‑noise ratio) support a smaller $d$. In our study, the data come from the numerical solution of equations of motion, which are inherently noiseless. We therefore set $d$ to a small value to ensure the precision of the LE computation.

In the second step, the two points $(\Omega(t_0), \Omega({t_0}'))$ evolve in steps $k$ along their trajectories. The evolved points are $(\Omega(t_0+k), \Omega({t_0}'+k))$, and the distance between them becomes ${L_0}' = ||\Omega(t_0+k)-\Omega({t_0}'+k)||$. This evolution, which leads to an increase in the exponent, is quantified by the following relation:
\begin{equation}
   \Delta\lambda = \frac{1}{k\cdot\Delta t}\log_{2}\frac{|\delta\mathbf{Z}(t)|}{|\delta\mathbf{Z}(0)|},
\end{equation}
where the prefactor $\frac{1}{k\cdot\Delta t}$ serves as a normalization constant and $\Delta t$ denotes the sampling interval between consecutive points in the time series.

After evolution, the separation ${L_0}'$ is compared with the prescribed radius $d$. If ${L_0}' \leq d$, the evolved pair is retained and the procedure proceeds to the next segment. If ${L_0}' > d$, a replacement neighbour must be chosen. We select a new point $\Omega({t_1}')$ in the neighbour of the current fiducial point $\Omega(t_0+k)$ such that the new separation
\begin{equation}
L_1 = ||\Omega(t_0+k) - \Omega({t_1}')||,
\end{equation}
satisfies $L_1 < d$. Furthermore, the direction of the new vector $\Omega(t_0+k)\Omega({t_1}')$ should be as close as possible to the old one $\Omega(t_0+k) \Omega({t_0}'+k)$. This process is equivalent to perform a Gram-Schmidt reorthonormalization in the tangent space of a continuous dynamical system. It ensures that we are tracing the most unstable direction, the one associated with the maximal LE. Finally, we repeat the second and third steps until the whole time series is traversed. The maximal LE is the time average of these local increasing rates:
\begin{equation}
   \lambda_{max} \approx \frac{1}{K} \sum^{K}_{i=1} \frac{1}{k_{i}\cdot\Delta t}\log_{2}\frac{{L_i}'}{L_i},
\end{equation}
where $K$ is the total number of segments (evolution-renormalization cycles) processed.

\subsection{Indicators of Chaotic Dynamics}\label{sub-ICD}

To characterize the chaotic dynamics we examine three complementary diagnostics: the reconstructed Lorenz attractor, the Poincaré section, and the maximal LE. Two upgrades are applied with respect to a minimal presentation. First, the Poincaré sections are computed from a densely sampled integration (about 120 radial periods, roughly $2.4$ times the span used elsewhere in this work), so that each section contains $190$--$320$ exact crossings of the equatorial plane; the points are drawn enlarged for print legibility. Second, the LE curves are shown separately for each spin, which turns out to be essential: the response of $\lambda_{\max}$ to the hair depends strongly on $S$.

\begin{figure}[htbp]
\centering
\includegraphics[scale=0.5]{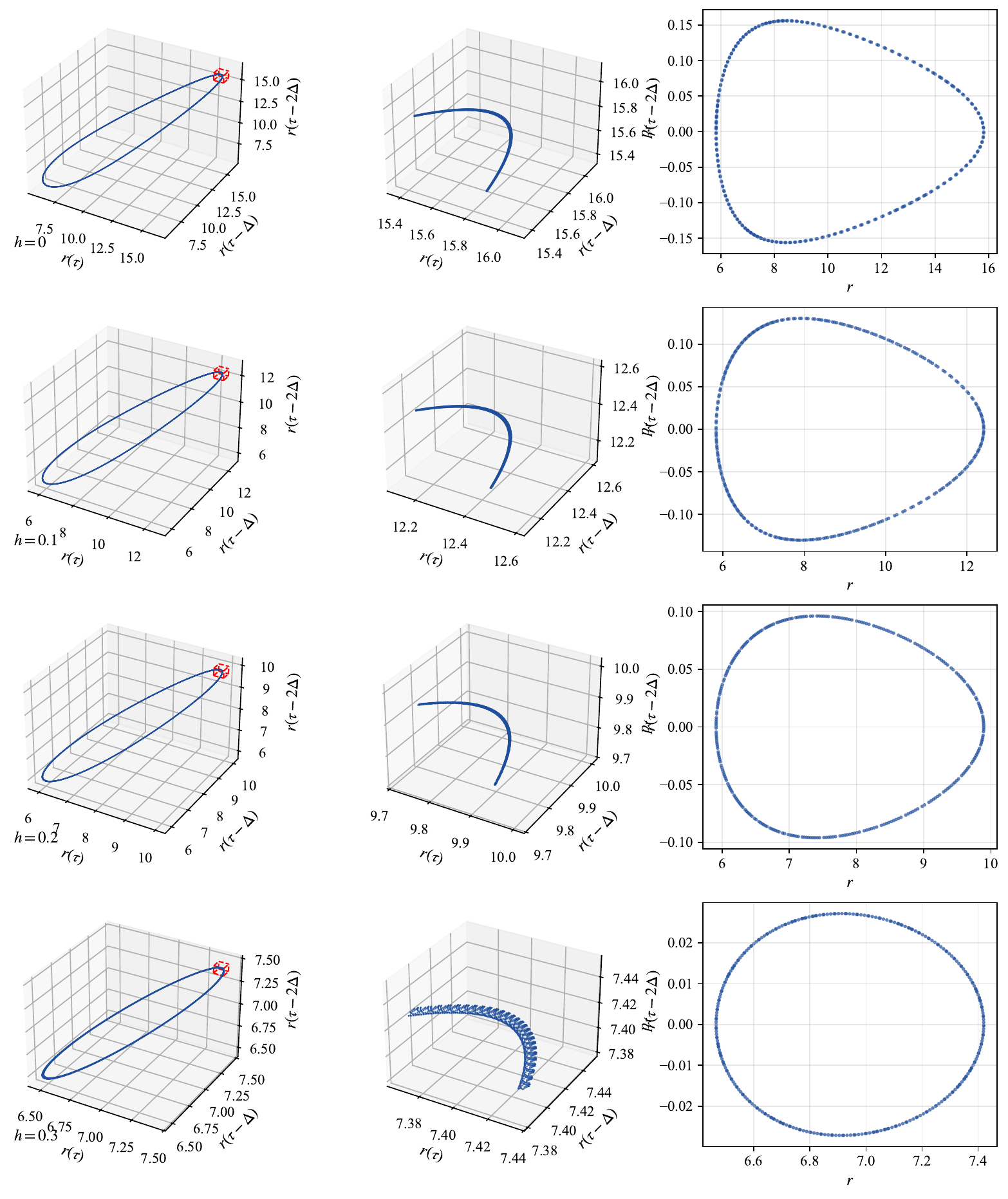}
\caption{Reconstructed attractors (left column; the red dashed box marks the region magnified in the middle column), magnified views of the boxed regions (middle column), and dense Poincaré sections (right column) for $S=0.5$ and $h=0$, $0.1$, $0.2$, $0.3$ (top to bottom).}
\label{fig:P&L}
\end{figure}

\begin{figure}[htbp]
\centering
\includegraphics[scale=0.5]{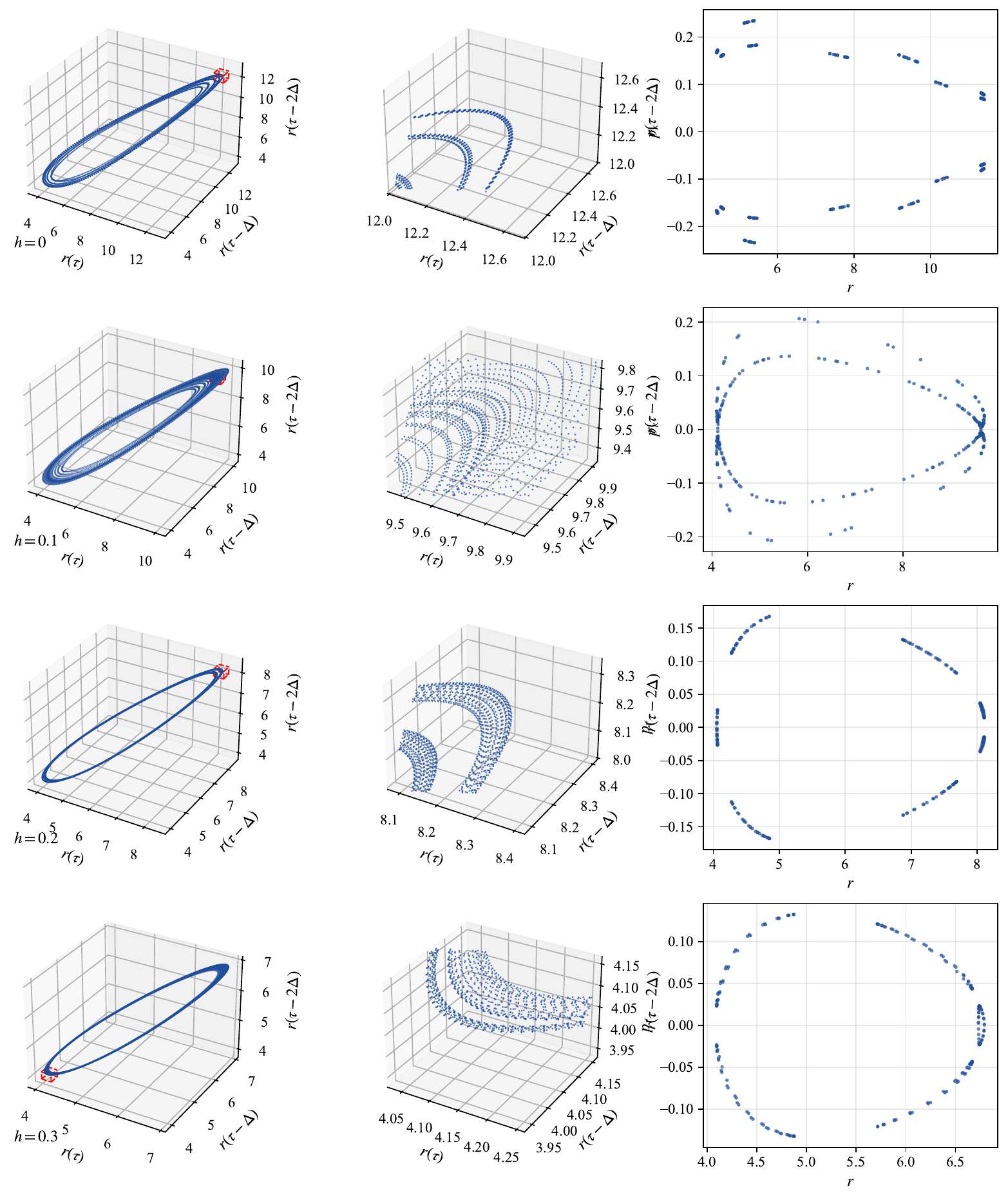}
\caption{Reconstructed attractors (left column; the red dashed box marks the region magnified in the middle column), magnified views of the boxed regions (middle column), and dense Poincaré sections (right column) for $S=1$ and $h=0$, $0.1$, $0.2$, $0.3$ (top to bottom).}
\label{fig:P&L_S1}
\end{figure}

Figure~\ref{fig:P&L} shows the results for $S=0.5$ and Fig.~\ref{fig:P&L_S1} for $S=1$, with rows corresponding to $h=0,0.1,0.2,0.3$. In each row, from left to right: the attractor reconstructed from the radial coordinate via Takens embedding, with its densest region marked by the red dashed box; the magnified view of the boxed region at its own scale; and the dense Poincaré section.

\begin{figure}[htbp]
\centering
\includegraphics[scale=0.72]{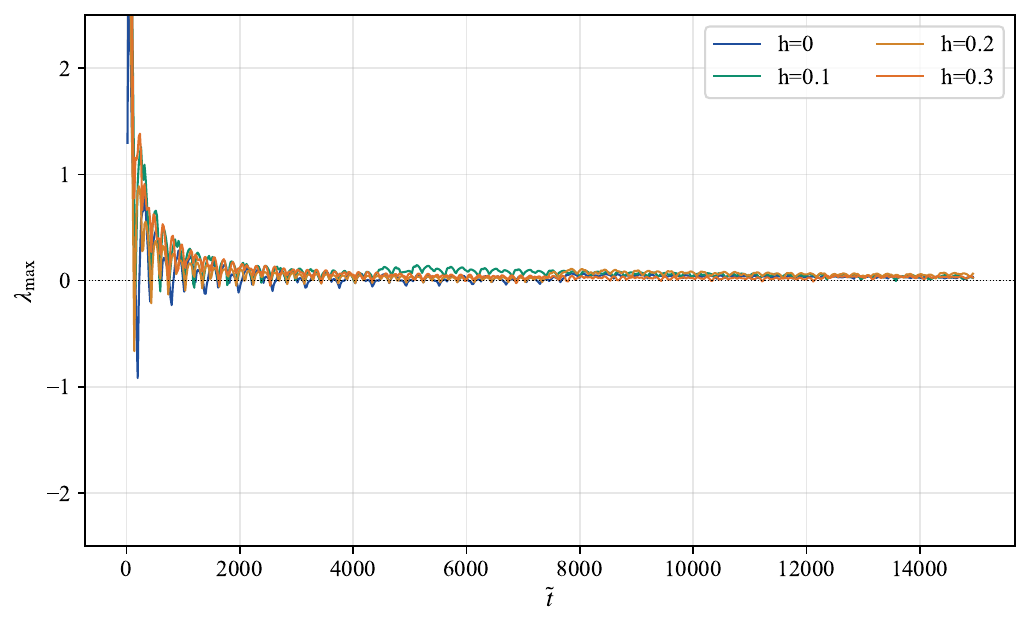}
\caption{Running maximal Lyapunov exponent versus $\tilde{t}$ for $S=0.5$ and $h=0$, $0.1$, $0.2$, $0.3$. The late-time block-averaged values are $\lambda_{\max}=0.035$, $0.025$, $0.066$, $0.028$.}
\label{fig:LE_S05}
\end{figure}

For $S=0.5$ (Fig.~\ref{fig:P&L}) the sections are thin closed bands—the signature of near-regular motion—with a thickness that grows mildly with $h$, while the attractor remains a well-defined, only slightly inflated loop. This is the regular regime (type B1 potential): the LE curves of Fig.~\ref{fig:LE_S05} stay below $\lambda_{\max}\simeq0.07$ throughout. For $S=1$ (Fig.~\ref{fig:P&L_S1}) the sections thicken into broad, partially filled bands and the attractor develops a visibly shattered, multi-strand structure: the orbit repeatedly grazes the saddle region, and neighbouring trajectories separate exponentially. This is the type B2 chaotic regime.

\begin{figure}[htbp]
\centering
\includegraphics[scale=0.72]{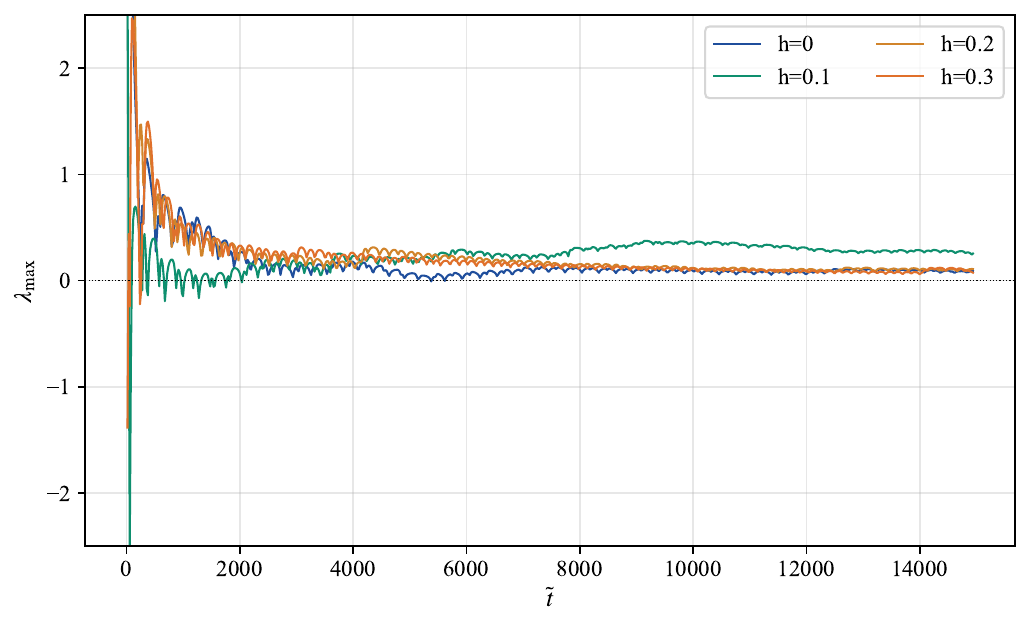}
\caption{Same as Fig.~\ref{fig:LE_S05} but for $S=1$: $\lambda_{\max}=0.090$, $0.255$, $0.109$, $0.075$—a strongly non-monotonic dependence on the hair, with its maximum at $h=0.1$.}
\label{fig:LE_S1}
\end{figure}

\begin{figure}[htbp]
\centering
\includegraphics[scale=0.72]{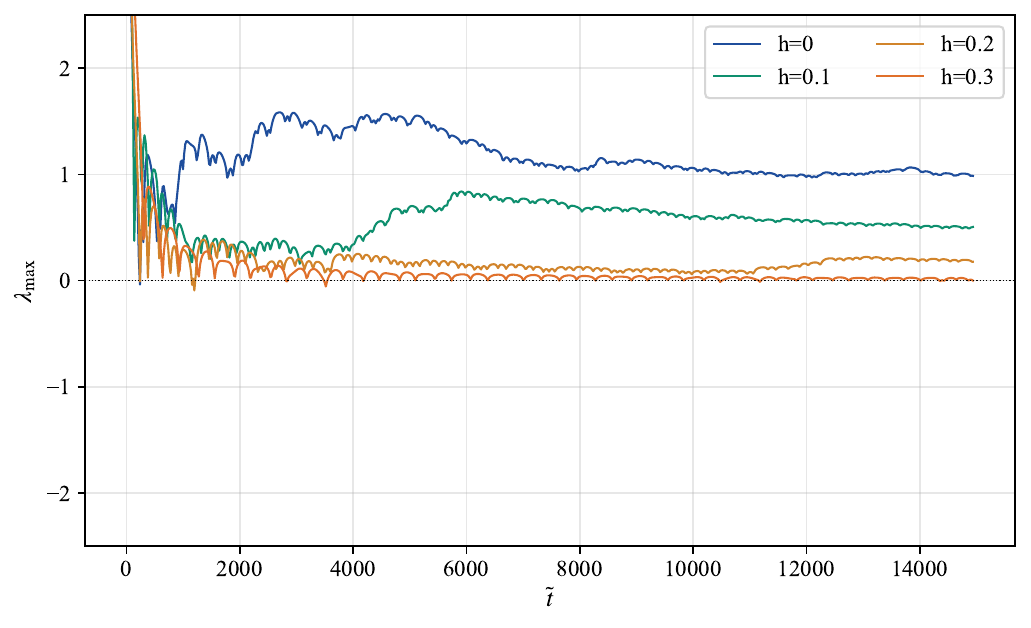}
\caption{Same as Fig.~\ref{fig:LE_S05} but for $S=1.5$: $\lambda_{\max}=0.986$, $0.505$, $0.178$, $0.001$—the only spin in our grid for which the Lyapunov exponent decreases monotonically with $h$.}
\label{fig:LE_S15}
\end{figure}

The maximal LEs are shown in Figs.~\ref{fig:LE_S05}, \ref{fig:LE_S1} and \ref{fig:LE_S15} for $S=0.5$, $1$ and $1.5$, respectively. The running exponent is plotted as a function of $\tilde{t}$; its late-time block-averaged value is read as the converged $\lambda_{\max}$. Three observations stand out. (i)~For $S=0.5$ the exponent remains small for all $h$, with a slightly non-monotonic variation ($\lambda_{\max}=0.035$, $0.025$, $0.066$, $0.028$ at $h=0,0.1,0.2,0.3$). (ii)~For $S=1$ the response is strongly non-monotonic: $\lambda_{\max}$ jumps from $0.090$ at $h=0$ to $0.255$ at $h=0.1$, drops to $0.109$ at $h=0.2$ and $0.075$ at $h=0.3$—the hair first amplifies the chaos by more than a factor of two and only then weakens it. (iii)~For $S=1.5$ alone the response is monotonic: $\lambda_{\max}$ falls steadily from $0.986$ at $h=0$ to $0.001$ at $h=0.3$, an almost complete quenching of the chaos.

The non-monotonic behavior at $S=0.5$ and $S=1$ is not a contradiction of the monotone erosion of the potential: the geometric measures of the saddle structure do decrease monotonically with $h$. Rather, the measured exponent results from the interplay between this monotone geometry and the resonance structure of the orbit. This is analyzed in the next subsection; the full quantitative details are deferred to Appendix~\ref{app-res}.

\subsection{Resonance Effects}\label{sub-res}

The peak of $\lambda_{\max}(h)$ at $S=1$, $h=0.1$ coincides with a low-order resonance of the orbital motion. The frequency ratios of Table~\ref{tab:resonance1} (Appendix~\ref{app-res}) tell a consistent story: at $h=0.1$ the azimuthal precession ratio has approached the $2{:}1$ commensurability ($\nu_\phi=1.952$) while the $1{:}1$ polar lock is broken ($\nu_\theta=1.055$), so the two lowest-order resonances overlap in the sense of Chirikov~\cite{Chirikov:1979zz} and their merged stochastic layers amplify the chaos; at $h=0.2$ the precession has crossed the $2{:}1$ line ($\nu_\phi=2.068$), leaving the orbit between low-order resonances, and $\lambda_{\max}$ falls back to $0.109$. The angular Poincaré sections of Appendix~\ref{app-res} display exactly this sequence: organized random layers and island-chains at $h=0$ and $h=0.2$;  at $h=0.1$, a featureless filled band provide a direct visualization of the mechanism.

\section{Gravitational Waveform}\label{sec-4}

In this section, we analyze the GW signals generated by the quadrupole moment of the orbit, focusing on the degree of match between its two polarization modes.

\subsection{Quadrupole Formula}\label{susec-QF}

To compute the gravitational waveform emitted by the spinning particle, we adopt the quadrupole formula, a well-established approximation valid for distant observers when the observer’s distance $r_{obs}$ greatly exceeds the orbital scale. This approach requires constructing the metric perturbation in the transverse-traceless (TT) gauge~\cite{poisson2014gravity,maggiore2008gravitational,Suzuki:1999si}. Below, we detail the key steps for calculating the two independent polarization modes $h_{+}$ and $h_{\times}$ of the GW, starting from the definition of the TT gauge and the mass-quadrupole moment.

We first set up an orthonormal triad aligned with the observer’s direction:
\begin{subequations}
    \begin{align}
        \mathbf{n} &= (\sin{\theta_{obs}} \cos{\phi_{obs}}, \sin{\theta_{obs}}\sin{\phi_{obs}}, \cos{\theta_{obs}}), \\
        \mathbf{m} &= (\cos{\theta_{obs}\cos{\phi_{obs}}}, \cos{\theta_{obs}}\sin{\phi_{obs}}, -\sin{\theta_{obs}}), \\
        \mathbf{l} &= (-\sin{\phi_{obs}}, \cos{\phi_{obs}}, 0),
    \end{align}
\end{subequations}
where the unit vector $\mathbf{n}$ points in the observer's direction, $\mathbf{m}$ and $\mathbf{l}$ form an orthogonal basis in the vertical plane to $\mathbf{n}$. Here, $\theta_{obs}$ and $\phi_{obs}$ denote the observer's polar and azimuthal angles, respectively. The corresponding “plus” and “cross” polarization tensors are then
\begin{subequations}
    \begin{align}
        ({e_{+}})_{ij} &= m_{i}m_{j} - l_{i}l_{j}, \\
        ({e_{\times}})_{ij} &= m_{i}l_{j} + l_{i}m_{j}.
    \end{align}
\end{subequations}

The second time‑derivative of the mass‑quadrupole moment in Cartesian coordinates is
\begin{equation}
   \ddot{Q}_{ij} = m(\ddot{x_{i}}x_{j} + 2\dot{x_{i}}\dot{x_{j}} + x_{i}\ddot{x_{j}}),
\end{equation}
where $x_i(t)$ are the coordinates obtained from integrating the MPD equations. To enforce the traceless condition we subtract the trace part of $\ddot{Q}_{ij}$:
\begin{equation}
   {\ddot{Q}_{ij}}^{tf} = \ddot{Q}_{ij} - \frac{1}{3}\delta_{ij}\sum_{k}\ddot{Q}_{kk}.
\end{equation}
Then we project the trace-free quadrupole moment with a TT projection operator:
\begin{equation}
   \Lambda_{ij,kl} = P_{ik}P_{jl} - \frac{1}{2}P_{ij}P_{kl},
\end{equation}
where the transverse projection operator $P_{ij}$ is given by
\begin{equation}
   P_{ij} = \delta_{ij} - n_i n_j.
\end{equation}
This leads to the expression for the metric perturbation:
\begin{equation}
   {h_{ij}} = \frac{2}{r_{obs}}\Lambda_{ij,kl}{\ddot{Q}_{kl}}^{tf}.
\end{equation}

The two independent polarization modes $h_{+}$ and $h_{\times}$ are obtained by contracting ${h_{ij}}$ the respective polarization tensors $({e_{+}})^{ij}$ and $({e_{\times}})^{ij}$, which yield:
\begin{subequations}
    \begin{align}
        \frac{1}{2}h_{+} \nonumber&= {h_{ij}} ({e_{+}})^{ij} \\
              \nonumber&= \left(h_{xx}-h_{yy}\right) \frac{\left(\cos^{2} \theta_{obs}+1\right)}{4} \cos 2 \phi_{obs} \\
              \nonumber& -\frac{\left(h_{xx}+h_{yy}-2 h_{zz}\right)}{4} \sin ^{2} \theta_{obs}+h_{xy}\left(\frac{\cos^{2} \theta_{obs}+1}{2}\right) \sin 2 \phi_{obs} \\
              & -h_{xz} \sin \theta_{obs} \cos \theta_{obs} \cos \phi_{obs}-h_{yz} \sin \theta_{obs} \cos \theta_{obs} \sin \phi_{obs}, \\
        \frac{1}{2}h_{\times} \nonumber&= h_{ij} ({e_{\times}})^{ij} \\
                   \nonumber&= -\frac{\left(h_{xx}-h_{yy}\right)}{2} \cos \theta_{obs} \sin 2 \phi_{obs}+h_{xy} \cos \theta_{obs} \cos 2 \phi_{obs} \\
                   & +h_{xz} \sin \theta_{obs} \sin \phi_{obs}-h_{yz} \sin \theta_{obs} \cos \phi_{obs}.
    \end{align}
\end{subequations}
In the following, we set the observer at the position $(r_{obs}, \theta_{obs}, \phi_{obs})=(5000, \frac{\pi}{4}, 0)$.

\subsection{Gravitational Waveform Signatures}\label{sub-waveform}

\begin{figure}[htbp]
	\subfigure[\ $h=0$]{
		\begin{minipage}{1\linewidth}
            \includegraphics[scale=0.525]{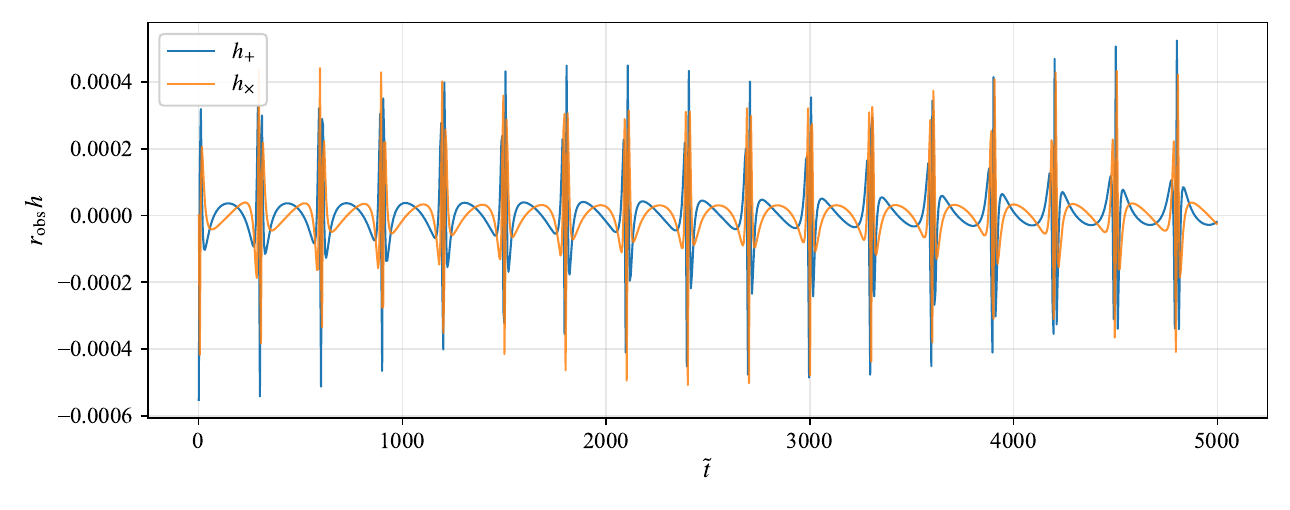}
		\end{minipage}
		}
    \subfigure[\ $h=0.1$]{
		\begin{minipage}{1\linewidth}
            \includegraphics[scale=0.525]{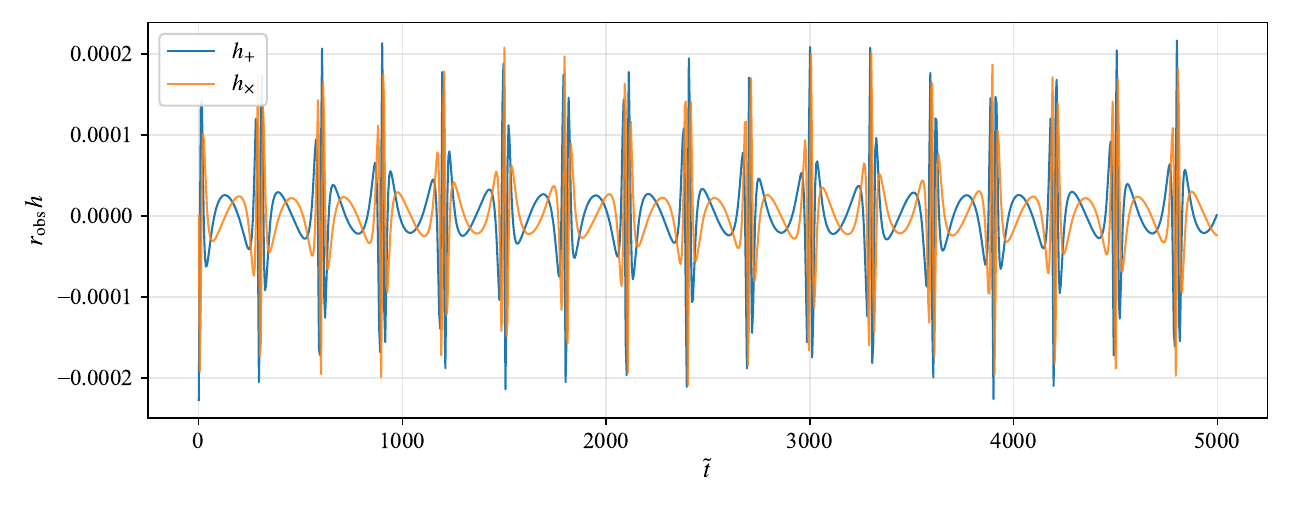}
		\end{minipage}
	}
    \subfigure[\ $h=0.2$]{
		\begin{minipage}{1\linewidth}
            \includegraphics[scale=0.525]{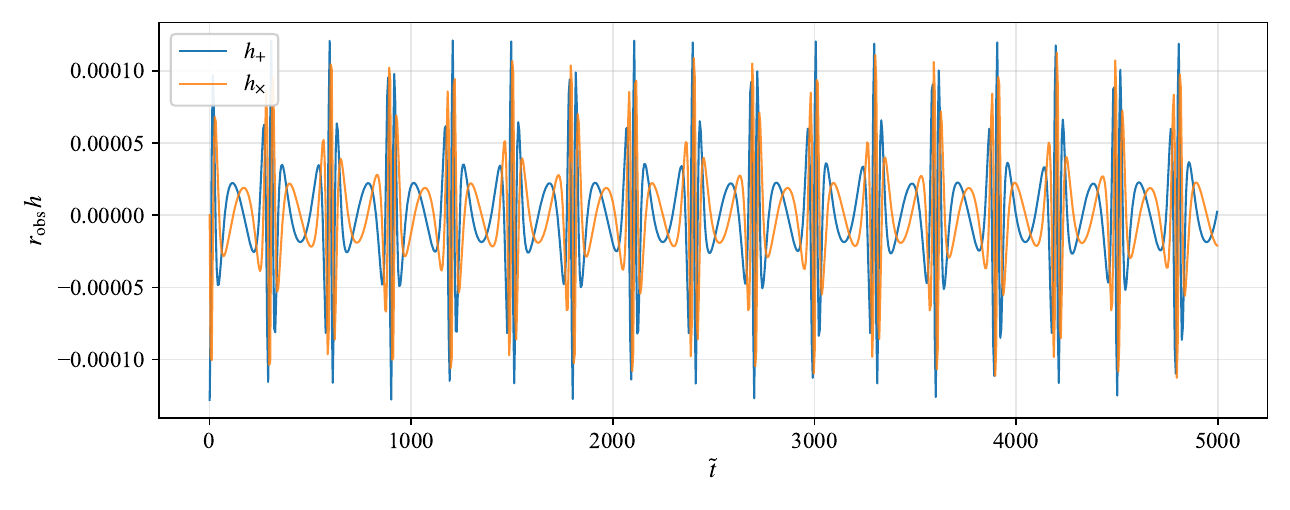}
		\end{minipage}
	}
    \subfigure[\ $h=0.3$]{
		\begin{minipage}{1\linewidth}
            \includegraphics[scale=0.525]{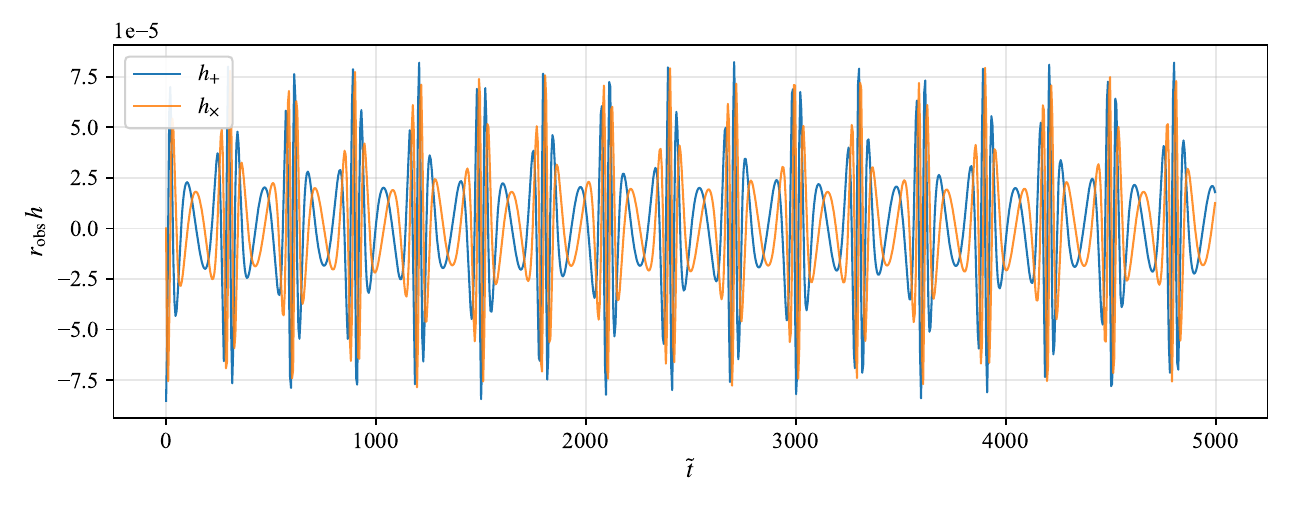}
		\end{minipage}
	}
	\caption{Gravitational waveforms for $S=0$ and $h=0$, $0.1$, $0.2$, $0.3$. The blue and orange lines represent the plus polarization $h_{+}$ and the cross polarization $h_{\times}$, respectively. Each panel shows a fixed window of the first 5000 samples.}
    \label{fig:waveform(S=0)}
\end{figure}
\begin{figure}[htbp]
	\subfigure[\ $h=0$]{
		\begin{minipage}{1\linewidth}
            \includegraphics[scale=0.525]{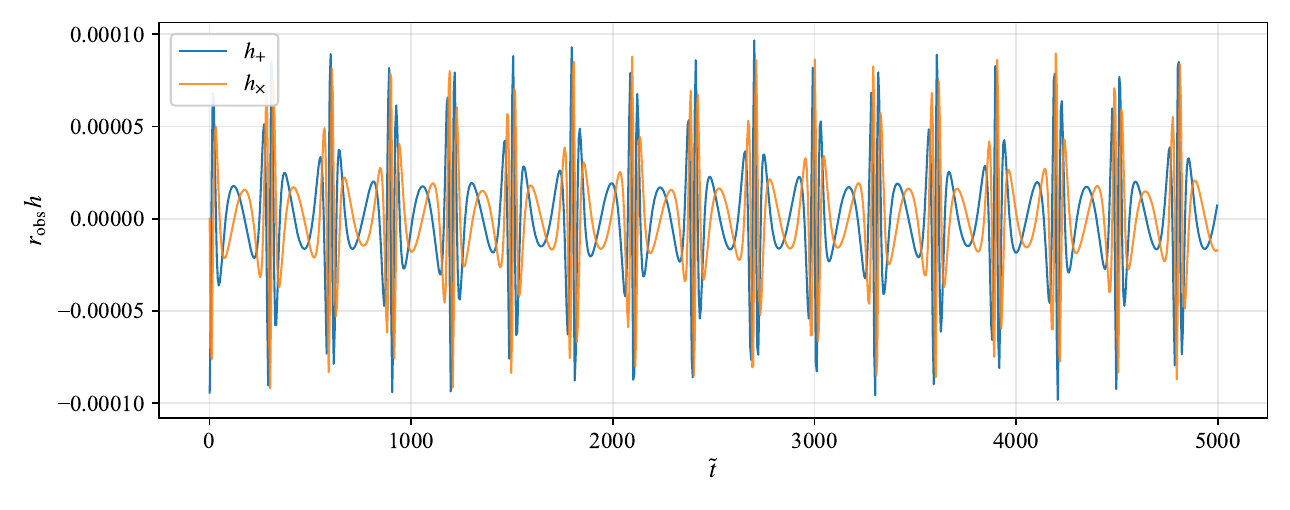}
		\end{minipage}
		}
    \subfigure[\ $h=0.1$]{
		\begin{minipage}{1\linewidth}
            \includegraphics[scale=0.525]{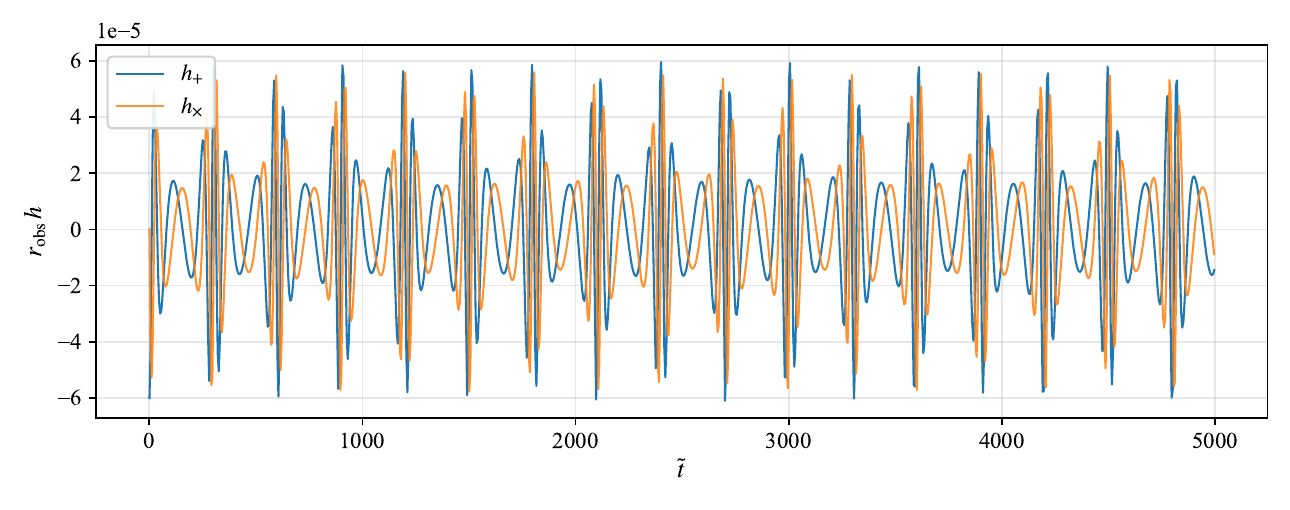}
		\end{minipage}
	}
    \subfigure[\ $h=0.2$]{
		\begin{minipage}{1\linewidth}
            \includegraphics[scale=0.525]{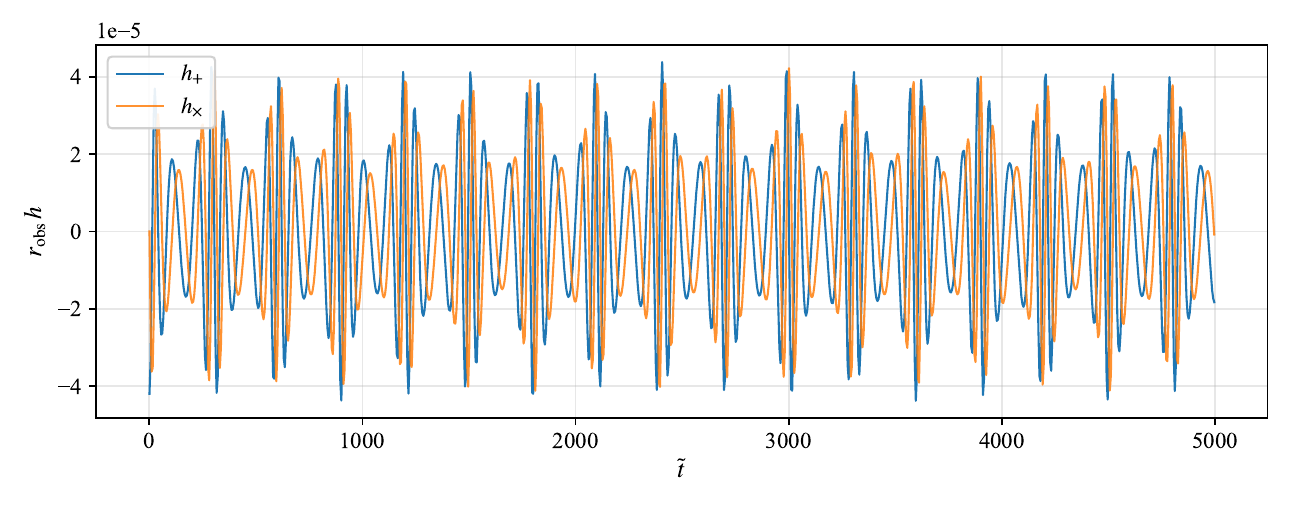}
		\end{minipage}
	}
    \subfigure[\ $h=0.3$]{
		\begin{minipage}{1\linewidth}
            \includegraphics[scale=0.525]{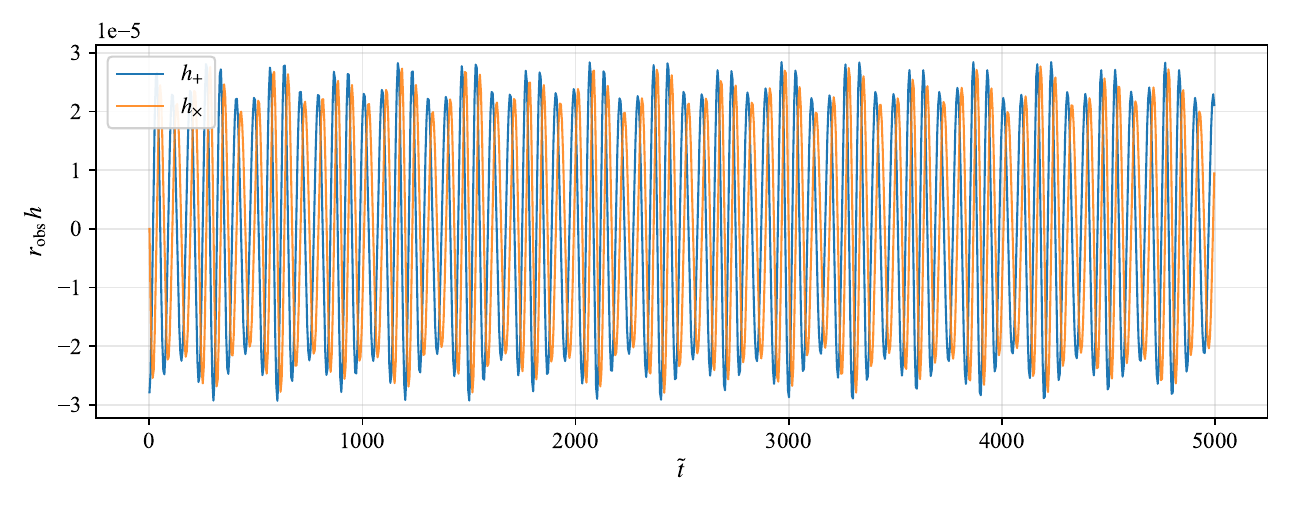}
		\end{minipage}
	}
	\caption{Gravitational waveforms for $S=0.5$ and $h=0$, $0.1$, $0.2$, $0.3$. The blue and orange lines represent the plus polarization $h_{+}$ and the cross polarization $h_{\times}$, respectively. Each panel shows a fixed window of the first 5000 samples.}
    \label{fig:waveform(S=0.5)}
\end{figure}
\begin{figure}[htbp]
	\subfigure[\ $h=0$]{
		\begin{minipage}{1\linewidth}
            \includegraphics[scale=0.525]{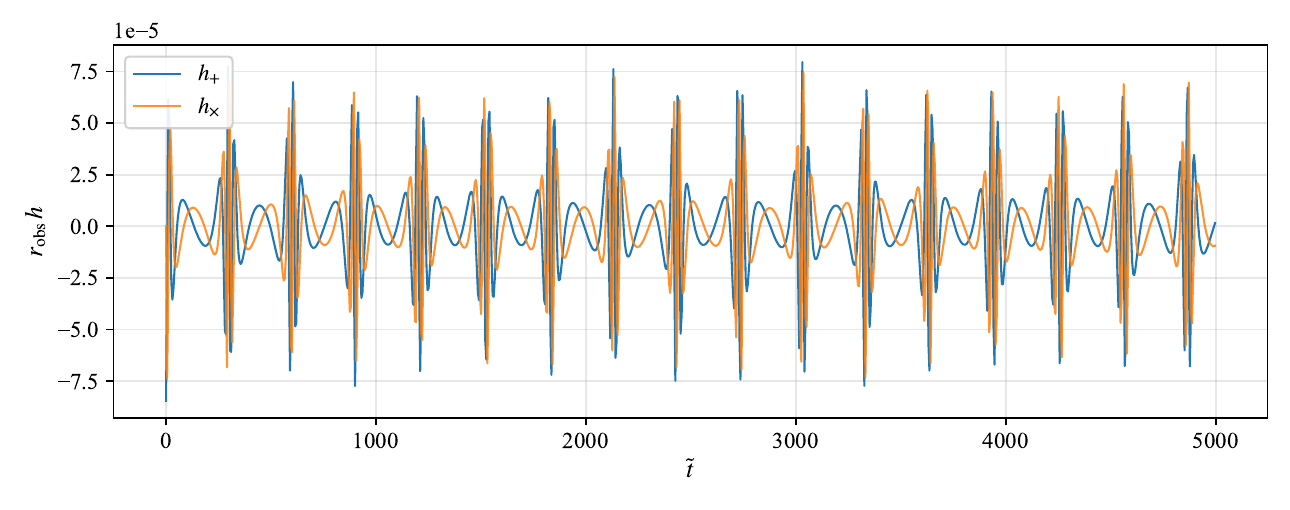}
		\end{minipage}
		}
    \subfigure[\ $h=0.1$]{
		\begin{minipage}{1\linewidth}
            \includegraphics[scale=0.525]{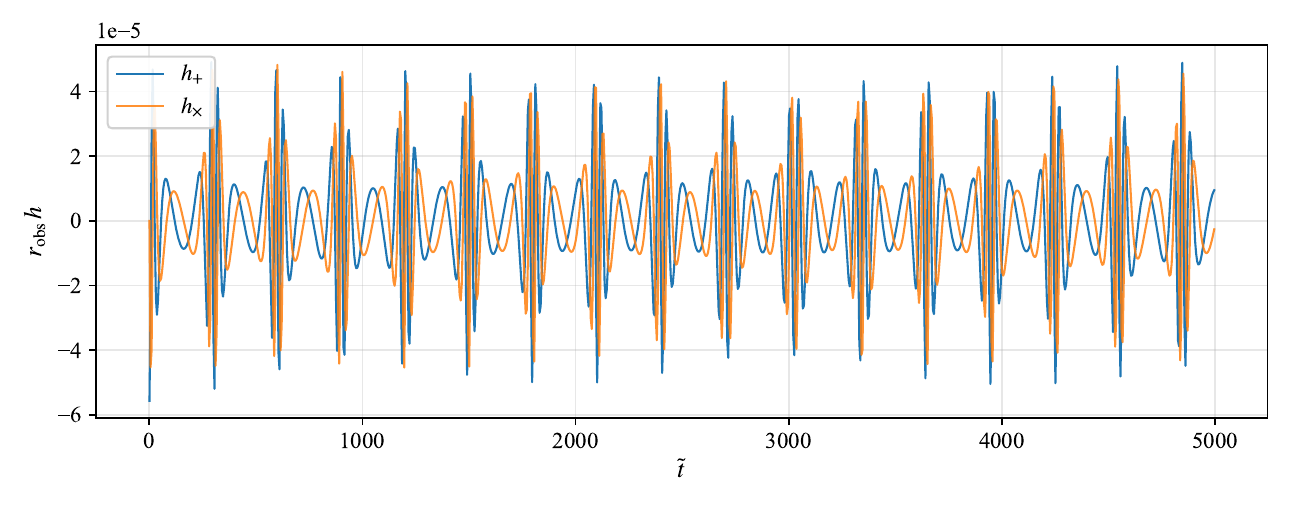}
		\end{minipage}
	}
    \subfigure[\ $h=0.2$]{
		\begin{minipage}{1\linewidth}
            \includegraphics[scale=0.525]{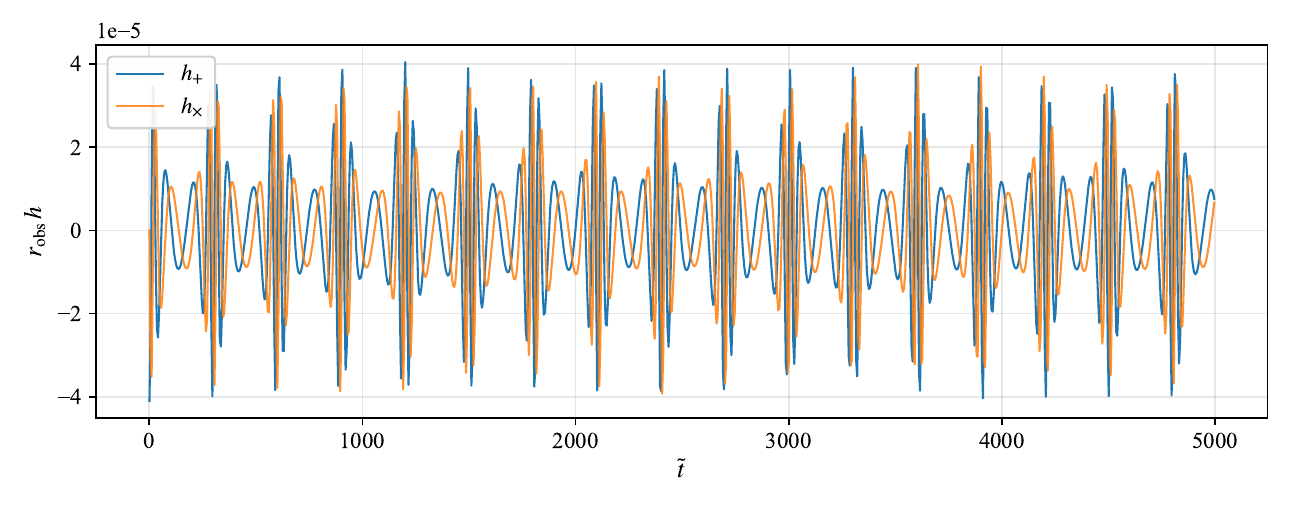}
		\end{minipage}
	}
    \subfigure[\ $h=0.3$]{
		\begin{minipage}{1\linewidth}
            \includegraphics[scale=0.525]{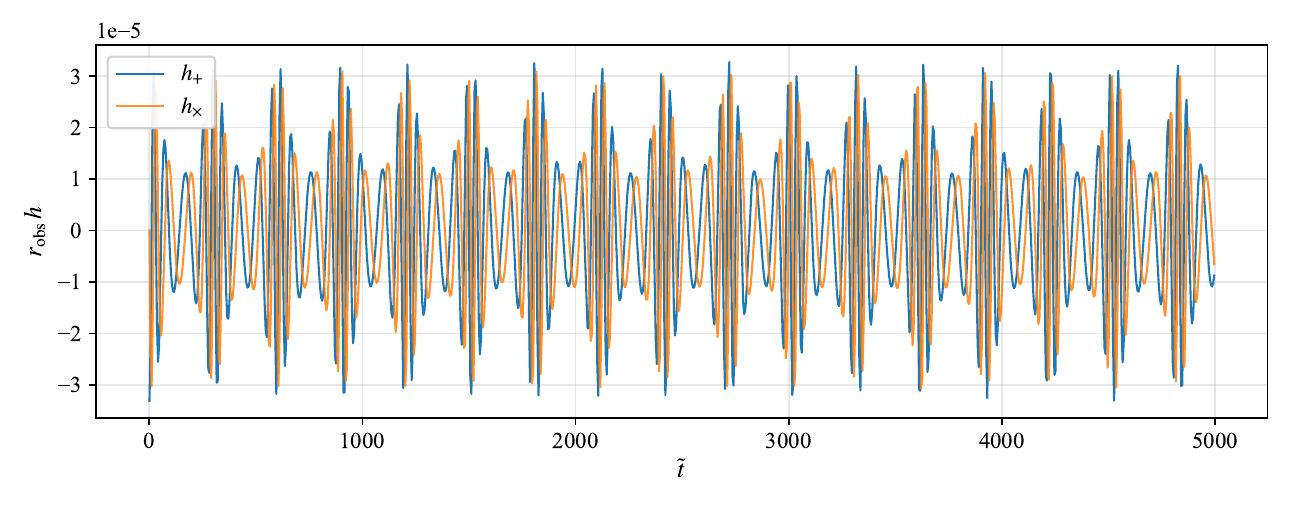}
		\end{minipage}
	}
	\caption{Gravitational waveforms for $S=1$ and $h=0$, $0.1$, $0.2$, $0.3$. The blue and orange lines represent the plus polarization $h_{+}$ and the cross polarization $h_{\times}$, respectively. Each panel shows a fixed window of the first 5000 samples.}
    \label{fig:waveform(S=1)}
\end{figure}

Before proceeding, we first introduce the key statistical metrics: the cross‑correlation (CC) coefficient and the mean absolute error (MAE) between $h_{+}$ and $h_{\times}$~\cite{Keita:2024skh} to quantify the evolving phase coherence. They are defined as
\begin{subequations}
    \begin{align}
    \mathrm{CC} &= \frac{\sum^{n}_{i=1}({h_{+}}^{i}-\bar{h}_{+})({h_{\times}}^{i}-\bar{h}_{\times})}{\sqrt{\sum^{n}_{i=1}({h_{+}}^{i}-\bar{h}_{+})^2\sum^{n}_{i=1}({h_{\times}}^{i}-\bar{h}_{\times})^2}},\label{eq:CC} \\
    \mathrm{MAE} &= \frac{1}{n}\sum^{n}_{i=1}|{h_{+}}^{i}-{h_{\times}}^{i}|\label{eq:MAE},
    \end{align}
\end{subequations}
where $n$ is the total number of time samples, and $\bar{h}_{+}$, $\bar{h}_{\times}$ denote the mean values of the respective time series.

Building on the quadrupole formula derived in subsection \ref{susec-QF}, we analyze the gravitational waveforms emitted by the spinning particle, specifically the two independent polarization modes $h_{+}$ and $h_{\times}$, and quantify how scalar hair modulates their phase coherence. The waveforms for varying spin $S=0, 0.5, 1$ and scalar hair parameter $h=0, 0.1, 0.2, 0.3$ are presented in Figs.~\ref{fig:waveform(S=0)}, \ref{fig:waveform(S=0.5)}, and \ref{fig:waveform(S=1)}, with key statistical metrics summarized in Table \ref{tableS} for all four spin values, including the strongly chaotic $S=1.5$.

\begin{table}[h]
\centering
\begin{tabular}{|c|c|c|c|c|c|c|c|c|}
\hline
\quad & \multicolumn{2}{c|}{$S=0$} & \multicolumn{2}{c|}{$S=0.5$} & \multicolumn{2}{c|}{$S=1$} & \multicolumn{2}{c|}{$S=1.5$} \\
\hline
\quad $h$ \ \quad & \quad \ CC \quad \quad & \quad MAE \ \ \quad & \quad \ CC \quad \quad & \quad MAE \ \ \quad & \quad \ CC \quad \quad & \quad MAE \ \ \quad & \quad \ CC \quad \quad & \quad MAE \ \ \quad \\
\hline
0   & 0.8627 & 0.283 & 0.9754 & 0.083 & 0.9338 & 0.206 & 0.9162 & 0.347 \\
0.1 & 0.9484 & 0.064 & 0.9252 & 0.054 & 0.9127 & 0.242 & 0.9454 & 0.272 \\
0.2 & 0.9600 & 0.170 & 0.9668 & 0.089 & 0.9457 & 0.139 & 0.9639 & 0.072 \\
0.3 & 0.9501 & 0.074 & 0.9881 & 0.119 & 0.9413 & 0.061 & 0.9908 & 0.051 \\
\hline
\end{tabular}
\caption{CC coefficient and MAE between $h_{+}$ and $h_{\times}$, evaluated using Eqs.~\eqref{eq:CC} and~\eqref{eq:MAE}; shown for $S = 0, 0.5, 1, 1.5$ and $h=0, 0.1, 0.2, 0.3$. Waveform figures are shown for $S=0$, $0.5$ and $1$}.\label{tableS}
\end{table}

The waveforms mirror the dynamical classification of Sec.~\ref{sec-3}. For the spinless particle ($S=0$, Fig.~\ref{fig:waveform(S=0)}) the two polarization modes keep a canonical phase relationship—differing by a constant phase shift of $\frac{\pi}{2}$~\cite{maggiore2008gravitational}—and the growing precession frequency with $h$ appears as a denser packing of the peaks, consistent with the shortened radial period observed in Sec.~\ref{sub-orbitalD}. For $S=0.5$ (Fig.~\ref{fig:waveform(S=0.5)}) mild dephasing appears at $h=0$, and for $S=1$ (Fig.~\ref{fig:waveform(S=1)}) the peak alignment becomes visibly irregular—the orbital chaos of the type B2 regime imprints itself directly on the relative phase of the two modes.

The CC and MAE values of Table~\ref{tableS} quantify this behavior and reveal that it is again state-specific rather than monotonic in $h$. For the chaotic spin $S=1$ the response to the hair is clearly non-monotonic: the dephasing is strongest at $h=0.1$, where CC reaches its minimum $0.913$ and MAE its maximum $0.242$, and then partially recovers at $h=0.2$ and $0.3$ (MAE $=0.139$ and $0.061$). This is precisely the same non-monotonic pattern displayed by the LE of Fig.~\ref{fig:LE_S1}, and it admits the same resonance interpretation: the $h=0.1$ orbit is the one in which the resonances overlap and the chaos is strongest, so the two polarization modes decohere most; at larger $h$ the resonance grip loosens and the waveform partially re-synchronizes. The spin dependence at fixed $h$ is likewise state-specific: the near-regular orbits are the most coherent ($S=0.5$: CC $\geq0.925$ at all $h$), the eccentric spinless rosette of $S=0$ already shows sizable MAE at $h=0$ ($0.283$) from its strongly modulated amplitude envelope rather than from chaos, and the chaotic $S=1$ sits in between ($0.206$) and is the only family in which the hair first amplifies the dephasing before suppressing it: its $h=0.1$ MAE ($0.242$) exceeds the $h=0$ value ($0.206$), whereas for every other spin the $h=0.1$ dephasing stays below its $h=0$ level. The $S=1.5$ family completes the picture: its coherence is restored monotonically with $h$ (CC rising from $0.916$ to $0.991$ and MAE falling from $0.347$ to $0.051$), in step with the monotonic quenching of its LE (Fig.~\ref{fig:LE_S15})---the only spin for which both the chaos and the waveform coherence respond monotonically to the hair. In line with the discussion of Sec.~\ref{sub-ICD}, we conclude that the GW polarization correlation tracks the state-specific, resonance-modulated chaos of the orbit rather than responding to the hair as a simple monotonic dial.


\section{Conclusions and Discussions}\label{conclusion}

We have studied the chaotic dynamics of a spinning test particle in the spacetime of a hairy BH within Horndeski gravity. By numerically integrating the MPD equations under the TD spin condition, with the angular momentum fixed and the energy tied to the structure of the effective potential, we analyzed the orbital morphology, attractors, Poincare sections, maximal LEs, and gravitational waveforms for spins $S=0$, $0.5$, $1$ and $1.5$ and hair parameters $h=0$--$0.3$.

Three results characterize the response of the system to the scalar hair. First, the hair determines which part of the parameter space is dynamically accessible at all: the bound region of the effective potential survives only up to a critical hair parameter $h_{\rm crit}(S)$, which is U-shaped in the spin (minimum $0.454$ near $S\simeq0.75$ at $J=4$), and beyond it no bound orbit exists. Second, the chaotic indicators do not respond to the hair as a simple monotonic suppression: at $S=0.5$ and $S=1$ the maximal LE varies non-monotonically with $h$---for $S=1$ it is amplified by almost a factor of three at $h=0.1$ before declining---while only at $S=1.5$ does it decrease monotonically, from $0.986$ to nearly zero. Third, the GW polarization correlation inherits the same state-specific behavior, with the strongest dephasing at $S=1$, $h=0.1$.

We have shown that the resonance structure of the orbit is the key ingredient behind this behavior, as the frequency ratios of Table~\ref{tab:resonance1} and the angular sections of Appendix~\ref{app-res} make visible: the peak of $\lambda_{\max}(h)$ at $S=1$ coincides with the overlap of the $1{:}1$ polar lock and the approaching $2{:}1$ precession resonance, and the dips with the loss of this overlap. The monotone erosion of the potential saddles by the hair proceeds in the background of this resonance dynamics; both must be considered together to understand how chaos around a hairy BH responds to the hair.

Our results caution against using any single chaotic indicator as a direct measure of the hair parameter: the same value of $\lambda_{\max}$ can belong to different combinations of $(S,h)$, and the sign of the response can change with the spin. What the chaos does provide is a sensitive, qualitatively rich fingerprint of the underlying potential geometry---through the existence range of bound orbits, the location of the response extrema, and the resonance content of the motion---which could be combined with waveform modeling to constrain deviations from GR in the strong-field regime.

Future work should extend this framework to rotating hairy BHs, where spin--spin interactions could further modulate the chaos and its resonances, and to more general Horndeski couplings. A systematic resonance mapping of the ($S,h$) parameter space, longer evolutions to sharpen the frequency and phase diagnostics, and waveform modeling that incorporates the resonance-modulated dephasing identified here will be essential for connecting these findings to future astrophysical observations.

\acknowledgments
We are especially grateful to Guoyang Fu for his valuable discussions. Yang Yu also wishes to thank Rui Bo and Qiaonan Li for their essential help with programming. This work is supported by the Natural Science Foundation of China under Grant Nos. 12375055, 12505085, the Jiangsu Postgraduate Research and Practice Innovation Program under Grant No. KYCX25$_{-}$3925, the China Postdoctoral Science Foundation (No. 2025T180931), and the Jiangsu Funding Program for Excellent Postdoctoral Talent (No. 2025ZB705).

\appendix

\section{Resonance effects and their quantification}\label{app-res}

This appendix documents the resonance analysis behind Sec.~\ref{sub-res}: the physical origin of the resonances, their frequency-ratio evidence, the direct visualization in the angular sections, and the reason why neither the effective potential nor the Poincar\'e section displays this resonance structure even though it shapes the LE.

\subsection{Origin of the resonances}

For an integrable spinless system the motion is quasi-periodic on an invariant torus, characterized by the fundamental frequencies of the radial, polar, and azimuthal modes~\cite{Lichtenberg:1992,Ott:1993}. When two frequency ratios approach a low-order rational number $p/q$, a $p{:}q$ resonance occurs: in the frame co-rotating with the combination frequency, the resonant phase
\begin{equation}
\Psi(t) = q\,\Phi_b(t) - p\,\Phi_a(t)
\end{equation}
(where $\Phi_{a,b}$ are the unwrapped phases of the two coupled modes) evolves slowly, executing a libration around the resonance center. In phase space, island chains of trapped trajectories form around the resonance, embedded in a stochastic layer along their separatrices. When two neighboring resonances overlap in the sense of Chirikov~\cite{Chirikov:1979zz}, the stochastic layers merge into a connected chaotic sea and the LE rises sharply. In our system the two relevant frequency ratios are the polar ratio $\nu_\theta=\omega_\theta/\omega_r$ (polar oscillations per radial period) and the azimuthal precession ratio $\nu_\phi$ (azimuthal advance per radial period); the coupling between the polar and azimuthal sectors is induced by the spin-curvature interaction, which is why the resonance dynamics is spin-dependent and strongest at intermediate spin.

\subsection{Frequency ratios and resonance proximity}

For each orbit we extract $\omega_r$ and $\omega_\theta$ from the peaks of the Hann-windowed fast Fourier transform (FFT) spectra of $r(\tau)$ and $\theta(\tau)$ (with parabolic peak interpolation), and $\nu_\phi$ from the total azimuthal advance per radial period. For the nearest low-order ($q\leq6$) commensurability $p/q$ of each ratio we record the detuning $\delta=|\nu-p/q|$; the ratios, their nearest commensurabilities and their detunings are collected in Table~\ref{tab:resonance1} and form the quantitative backbone of the resonance interpretation in the main text.

\begin{table}[h]
\centering
\begin{tabular}{|c|c|c|c|c|c|c|c|c|}
\hline
$S$ & $h$ & $\lambda_{\max}$ & $\nu_\theta$ & nearest & $\delta_\theta$ & $\nu_\phi$ & nearest & $\delta_\phi$ \\
\hline
0.5 & 0   & 0.035 & 1.246 & 5:4 & 0.004 & 1.638 & 5:3  & 0.029 \\
0.5 & 0.1 & 0.025 & 1.255 & 5:4 & 0.005 & 1.721 & 7:4  & 0.029 \\
0.5 & 0.2 & 0.066 & 1.281 & 5:4 & 0.031 & 1.837 & 11:6 & 0.004 \\
0.5 & 0.3 & 0.028 & 1.327 & 4:3 & 0.006 & 1.999 & 2:1  & 0.001 \\
1.0 & 0   & 0.090 & 1.000 & 1:1 & 0.000 & 1.887 & 11:6 & 0.054 \\
1.0 & 0.1 & 0.255 & 1.055 & 1:1 & 0.055 & 1.952 & 2:1  & 0.048 \\
1.0 & 0.2 & 0.109 & 0.874 & 5:6 & 0.041 & 2.068 & 2:1  & 0.068 \\
1.0 & 0.3 & 0.075 & 0.831 & 5:6 & 0.002 & 2.292 & 9:4  & 0.042 \\
1.5 & 0   & 0.986 & 0.663 & 2:3 & 0.004 & 1.627 & 8:5  & 0.027 \\
1.5 & 0.1 & 0.505 & 0.500 & 1:2 & 0.000 & 1.667 & 5:3  & 0.001 \\
1.5 & 0.2 & 0.178 & 0.381 & 2:5 & 0.019 & 1.476 & 3:2  & 0.024 \\
1.5 & 0.3 & 0.001 & 0.239 & 1:4 & 0.011 & 1.548 & 3:2  & 0.048 \\
\hline
\end{tabular}
\caption{Frequency ratios and resonance proximity for $S=0.5$, $1$, $1.5$. The $S=0$ orbits are exactly equatorial ($\theta\equiv\pi/2$) and admit no polar ratio.}\label{tab:resonance1}
\end{table}

Two entries of Table~\ref{tab:resonance1} deserve comment. At $S=0.5$, $h=0.3$ the orbit sits almost exactly on the $2{:}1$ precession commensurability ($\nu_\phi=1.999$, $\delta=0.001$) yet remains regular with a small $\lambda_{\max}$: resonance proximity alone does not imply chaos, in agreement with the standard resonance picture~\cite{Lichtenberg:1992}. Conversely, at $S=1.5$, $h=0.1$ the motion is strongly chaotic ($\lambda_{\max}=0.505$) although the polar ratio is exactly $1{:}2$: in the broadband regime a rational frequency ratio does not by itself imply a functional lock, which is why the angular sections displayed below---rather than detuning numbers alone---are the decisive evidence there. Finally, the near-vanishing $\lambda_{\max}=0.001$ at $S=1.5$, $h=0.3$ has yet another origin: the off-plane saddles have dived into the equatorial plane (Sec.~\ref{sub-eff-p}), the saddle--well gap has collapsed to $3\times10^{-4}$, and the chaotic neck has vanished altogether.

\begin{figure}[htbp]
\centering
\includegraphics[scale=0.5]{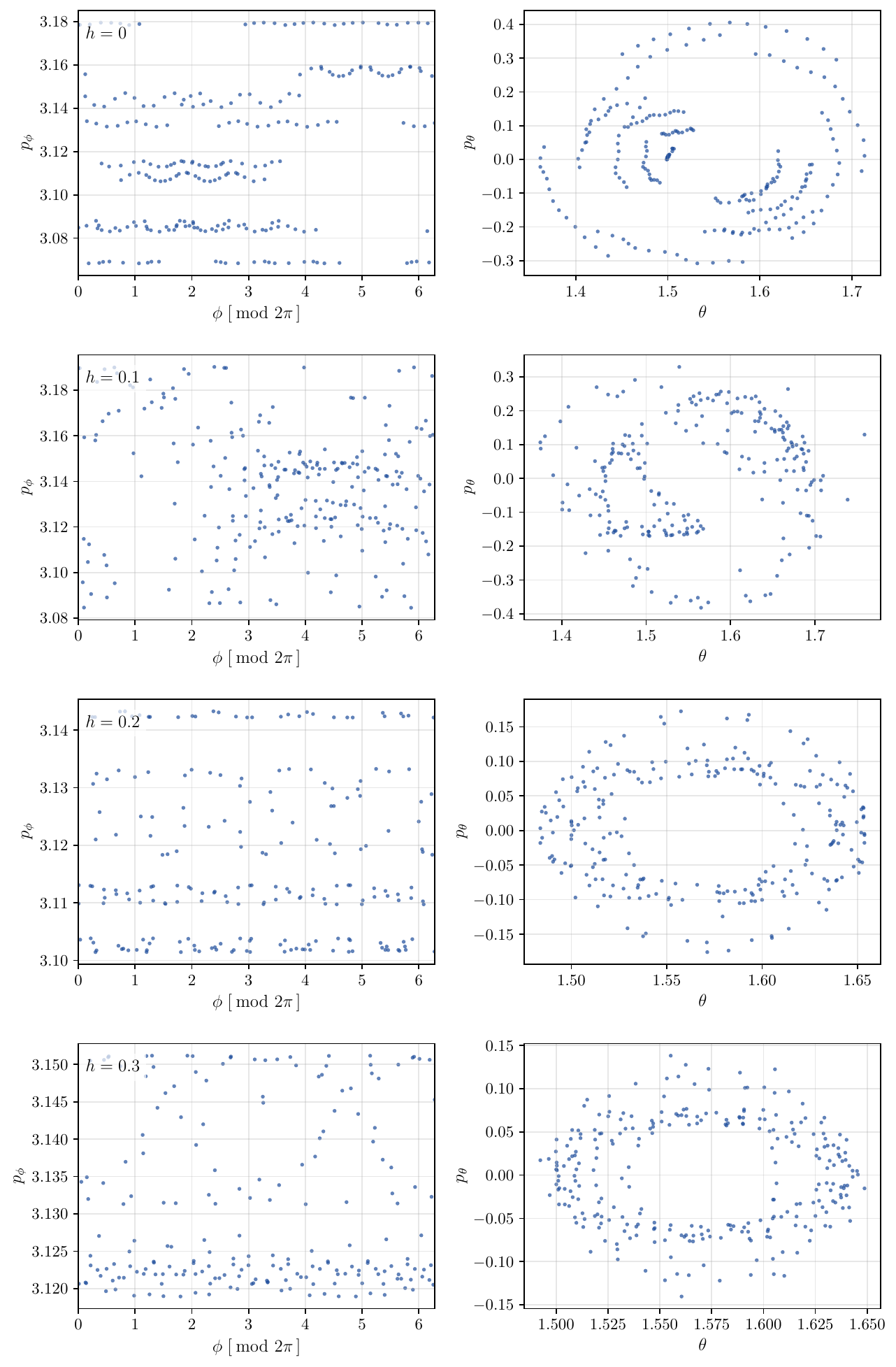}
\caption{Angular Poincar\'e sections of the $S=1$ orbits. Left column: $(\phi\,\bmod\,2\pi,\,p_\phi)$ sampled at the $\theta=\pi/2$ crossings. Right column: $(\theta,p_\theta)$ sampled at the $\phi=0$ crossings. Rows correspond to $h=0$, $0.1$, $0.2$, $0.3$ (top to bottom).}
\label{fig:ang-S1}
\end{figure}

The sections make the resonance mechanism of Sec.~\ref{sub-res} directly visible. At $h=0$, the $(\phi,p_\phi)$ section shows narrow horizontal bands (random layers)---the footprints of the polar-azimuthal resonance islands, with $p_\phi$ sampling discrete levels as the orbit precesses---and the $(\theta,p_\theta)$ section show structured arcs rather than a filled band. At $h=0.1$, the precession has been quite closed to the $2{:}1$ line and the polar lock is broken (Table~\ref{tab:resonance1}), and the bands and island-chains correspondingly disappears, while the $\lambda_{\max}$ arises to $0.255$. At $h=0.2$ the band structure partially re-emerges, matching the decline of $\lambda_{\max}$ to $0.109$. At $h=0.3$, the $\lambda_{\max}$ further shrinks to $0.075$. The sections thus provide the direct, visual counterpart of the frequency-ratio argument.

\subsection{Why the resonance is invisible in the effective potential and in the Poincar\'e section}

The effective potential $V_{-}(r,\theta)$ is static geometry determined by the conserved $(E,J)$: it answers where motion is allowed, but carries no information about at what tempo the orbit traverses the allowed region. The resonance condition is a Diophantine condition on integrated period ratios—quantities defined only after the equations of motion have been solved—so no contour of $V_{-}$ changes character when a frequency ratio crosses a rational value. Kolmogorov--Arnold--Moser (KAM) theory and the Poincar\'e-Birkhoff theorem likewise classify regions by torus winding numbers, not by potential values~\cite{Lichtenberg:1992,Ott:1993}. Numerically, the effect is dramatic: between $h=0$ and $h=0.1$ at $S=1$ the barrier height falls by only about $2\%$ and the saddle hyperbolicity by $10\%$, while $\nu_\phi$ moves from $1.887$ to $1.952$—across the Chirikov overlap threshold of the $2{:}1$ resonance—and $\lambda_{\max}$ jumps from $0.090$ to $0.255$.

The Poincar\'e section used in this work, $(r,p_r)$ at $\theta=\pi/2$ crossings, is invisible to the resonance for a complementary reason: the locking occurs between the two angular degrees of freedom (polar and azimuthal, coupled by the spin-orbit interaction), and the section projects both of them out. The lock changes how many crossings occur per radial period and when they occur, not where in $(r,p_r)$ they land: resonant and non-resonant orbits leave nearly identical annular footprints on the equatorial section. In the standard language, a polar-radial ($\theta{:}r$) resonance would appear as island chains in the $(r,p_r)$ surface~\cite{Lichtenberg:1992,Ott:1993}; the dominant locks of this system, however, are polar-azimuthal, which would only appear in a $(\phi,p_\phi)$ or $(\theta,p_\theta)$ section. Accordingly, the visual appearance of the equatorial sections is governed by the radial stochastic layer around the saddle necks; the resonance information must be read from the angular sections themselves. We have verified this prediction directly: Fig.~\ref{fig:ang-S1} above shows both angular sections of the $S=1$ orbit, and the sequence organized bands $\to$ merged chaotic sea $\to$ partial re-organization tracks the frequency ratios of Table~\ref{tab:resonance1} and the Lyapunov peaks of Fig.~\ref{fig:LE_S1}. Any diagnostic that projects out the resonating degree of freedom---be it the potential surface or the equatorial section---is blind to the resonance.

\bibliographystyle{utphys}
\bibliography{1Ref}

\end{document}